\documentclass[12pt,a4paper]{article}

\usepackage{epsfig}

\newcommand{\modii}{\mathrm{I\!I}}

\begin{document}
\vspace*{-3cm}
\begin{flushright}
FISIST/16--2002/CFIF \\
hep-ph/0210112 \\
October 2002
\end{flushright}
\vspace{0.5cm}
\begin{center}
\begin{Large}
{\bf Effects of mixing with quark singlets}
\end{Large}

\vspace{0.5cm}
J. A. Aguilar--Saavedra \\
{\it Departamento de F\'{\i}sica and Grupo de F\'{\i}sica de
Part\'{\i}culas (GFP), \\
Instituto Superior T\'ecnico, P-1049-001 Lisboa, Portugal}
\end{center}

\begin{abstract}
The mixing of the known quarks with new heavy singlets can modify significantly
some observables with respect to the Standard Model predictions. We analyse the
range of deviations permitted by the constraints from precision electroweak data
and flavour-changing neutral processes at low energies. We study top charged
current and neutral current couplings, which will be directly tested at top
factories, like LHC and TESLA. We discuss some examples of observables from $K$
and $B$ physics, as the branching ratio of $K_L \to \pi^0 \nu \bar \nu$, the
$B^0_s$ mass difference or the time-dependent CP asymmetry in
$B_s^0 \to D_s^+ D_s^-$, which can also show large new effects.
\end{abstract}

\section{Introduction}
The successful operation of LEP and SLD in the past few years has provided
precise experimental data \cite{papiro1,papiro1b} with which the Standard Model
(SM) and its proposed extensions must be confronted. The results for
$\varepsilon'/\varepsilon$ have converged \cite{papiro2a,papiro2b}, providing
evidence for direct CP violation in the neutral kaon system. In addition, $B$
factories have begun producing data on $B$ decays and CP violation, which test
the Cabibbo-Kobayashi-Maskawa (CKM) matrix elements involving the top quark and
the CKM phase $\delta$. However, the determination of most parameters involving
the top quark is still strongly model-dependent. While the CKM matrix elements
that mix light quarks are extracted from tree-level processes (and hence their
measurement is model-independent to a large extent), the charged current
couplings $V_{td}$ and $V_{ts}$ are derived from one-loop processes
\cite{papiro3}, to which new physics beyond the SM may well contribute. The
Tevatron determination of $V_{tb}$ in top pair production \cite{papiro4} is
obtained assuming $3 \times 3$ CKM unitarity, and the neutral current
interactions of the top with the $Z$ boson remain virtually unknown from the
experimental point of view. This fact contrasts with the high precision achieved
for the couplings of the $b$ and $c$ quarks at LEP and SLD, obtained from the
ratios $R_b$, $R_c$ and the forward-backward (FB) asymmetries
$A_\mathrm{FB}^{0,b}$, $A_\mathrm{FB}^{0,c}$.

The situation concerning CP violating phases is better (see for instance
Ref.~\cite{papiro5a,papiro5}). Few years ago,
the single phase $\delta$ present in the CKM matrix could merely be adjusted to
reproduce the experimental value of the only CP violation observable available,
$\varepsilon$ in the kaon system. With the resolution of the conflict between
the NA31 and E731 values of $\varepsilon'/\varepsilon$, and the recent
measurement of the CP asymmetry $a_{\psi K_S}$ in the $B$ system
\cite{papiro6a,papiro6b}, there are two new CP violation observables, both in
agreement with the SM predictions, which allow to test the CKM picture of CP
violation. Experiments under way at $B$ factories keep investigating other CP
asymmetries to dig out the phase structure of the CKM matrix. Likewise, the
knowledge of the top quark properties will improve in the next years, with the
arrival of top factories, LHC and TESLA \cite{papiro7,papiro8}. For instance,
single top production at LHC \cite{papiro9a,papiro9b,papiro9c} will yield a
measurement
of $V_{tb}$ with an accuracy of $\pm$ 7\%. In top pair production, the angular
distributions of the top decay products will provide a very precise
determination of the structure of the $Wtb$ vertex, even sensitive to QCD
corrections \cite{papiro10}. The prospects for $V_{td}$ and $V_{ts}$ are less
optimistic due to the difficulty in tagging the light quark jets.

Before top factories come into operation,  it is natural to ask ourselves how
large the departures from the SM predictions might be. Answering this question
means knowing how precisely one can indirectly fix  the allowed values of the
least known parameters, taking into account all the present relevant data from
electroweak precision measurements and from kaon, $D$ and $B$ physics. We will
show that there is still large room for new physics, which may manifest itself
in the form of deviations of the properties of the known quarks from SM
expectations. This is especially the case for the top quark, whose couplings are
poorly known, and also for rare $K$ decays and CP asymmetries in the $B$
systems, which are currently being probed at $K$ and $B$ factories.

With this aim we study a class of SM extensions in which $Q=2/3$ up-type or
$Q=-1/3$ down-type quark singlets are added to the three SM families
\cite{papiro11a,papiro11b,papiro11c,papiro11c2,papiro11d,papiro11e,papiro11f,
papiro11g,papiro11h,papiro11i,papiro11j}. These exotic quarks, often called
vector-like, have both their left and right components transforming as singlets
under $\mathrm{SU}(2)_L$, and therefore their addition to the SM quark content
does not spoil the cancellation of the triangle anomalies. In these models,
which are described in the next Section, $3 \times 3$ CKM unitarity does not
necessarily hold, and mixing of the new quarks with the standard ones can lead
to sizeable departures from the SM predictions
\cite{papiro12a,papiro12b,papiro12c,papiro12d}.
For instance, the CKM matrix elements $V_{td}$, $V_{ts}$ and $V_{tb}$ and the
top neutral current couplings with the $Z$ boson can  be quite different from SM
expectations. The ratio of branching fractions of the ``golden modes'' 
$\mathrm{Br}(K_L \to \pi^0 \nu \bar \nu) /
\mathrm{Br}(K^+ \to \pi^+ \nu \bar \nu)$ can have an enhancement of one order
of magnitude with respect to the SM prediction, and the time-dependent CP
asymmetry in the decay $B_s^0 \to D_s^+ D_s^-$, which is predicted to be very
small in the SM, can vary between $-0.4$ and $0.4$.

Apart from their simplicity and the potentially large effects on experimental
observables, there are several theoretical reasons to consider quark
isosinglets. Down singlets appear in grand unification theories
\cite{papiro11e,papiro13a,papiro13b}, for instance those based on the
gauge group $E_6$ (in the 27 representation of $E_6$ a $Q=-1/3$ singlet is
associated to each fermion family).
The presence of down singlets does not spoil gauge coupling unification, as
long as they are embedded within the 27 representation of $E_6$
\cite{papiro11h,papiro11i}. When added to the SM particle content, they can
improve the convergence of the couplings, but not as well as in the minimal
supersymmetric SM \cite{papiro14a}.
Models with large extra dimensions with for
instance $t_R$ in the bulk predict the existence of a tower of $Q=2/3$ singlets
$T_{L,R}^{(n)}$. If there is multilocalisation the lightest one,
$T_{L,R}^{(1)}$, can have a mass $m_T \sim 300$ GeV or larger and an observable
mixing with the top \cite{papiro14}. Similarly, if $b_R$ is in the bulk, there
exists a tower of $B_{L,R}^{(n)}$, but the mixing with SM fermions is suppressed
in relation to the top case if the Higgs is restricted to live on the boundary.

There are three recent studies regarding the constraints on models with extra
singlets \cite{papiro15a,papiro15b,papiro15c}. In this work, we extend these
analyses in three main aspects: ({\em i\/}) We include up singlets, as well as
down singlets, referring to them as Model I and $\modii$, respectively.
({\em ii\/}) We study the limits on $V_{td}$, $V_{ts}$, $V_{tb}$, top neutral
current couplings and other observables not previously analysed;
({\em iii\/}) We take a larger set of experimental constraints into account:
the correlated measurement of $R_b$, $R_c$, $A_\mathrm{FB}^{0,b}$,
$A_\mathrm{FB}^{0,c}$, $\mathcal{A}_b$, $\mathcal{A}_c$; oblique corrections;
the $|\delta m_B|$, $|\delta m_{B_s}|$ and $|\delta m_D|$ mass differences; the
CP violation observables $\varepsilon$, $\varepsilon'/\varepsilon$,
$a_{\psi K_S}$; the decays $b \to s \gamma$, $b \to s \mu^+ \mu^-$,
$b \to s e^+ e^-$, $K^+ \to \pi^+ \nu \bar \nu$ and $K_L \to \mu^+ \mu^-$;
$\nu N$ processes and atomic parity violation. In addition, we examine several
other potential restrictions, which  turn out to be less important than the
previous ones. We allow mixing of all the generations with either $Q=2/3$ or
$Q=-1/3$ exotic quarks, and we consider that one or two singlets can mix
significantly, though for brevity in the notation we always refer to one extra
singlet $T$ or $B$.

This paper is organised as follows: In Section~\ref{sec:2} the main features of
the models are described. In Section~\ref{sec:3} we summarise the direct limits
on CKM matrix elements and the masses of the new quarks. In Section~\ref{sec:4}
we review the constraints from precision electroweak data: $R_b$, $R_c$,
asymmetries and oblique corrections. In Section~\ref{sec:5} we focus our
attention on flavour-changing neutral (FCN) processes at low energies: meson
mixing, $B$ decays and kaon decays. The various constraints on the $Z$ couplings
of the $u$, $d$ quarks are studied in Section~\ref{sec:6}. We introduce the
formalism necessary for the discussion of some observables from $K$ and $B$
physics in Section~\ref{sec:7}. We present the results in Section~\ref{sec:8},
and in Section~\ref{sec:9} we draw our conclusions. In Appendix~\ref{sec:ap1}
we collect the common input parameters for our calculations, and in
Appendix~\ref{sec:ap2} the Inami-Lim functions needed. The statistical
prescriptions used in our analysis are explained in Appendix~\ref{sec:ap3}.

\section{Brief description of the models}
\label{sec:2}
In order to fix our notation briefly, in this Section we will be a little more
general than needed in the rest of the paper (see for instance
Ref.~\cite{papiro16} for an extended discussion including isodoublets and mirror
quarks too). We consider a SM extension with $N$ standard quark families and
$n_u$ up, $n_d$ down vector-like singlets. The total numbers of up and down
quarks, $\mathcal{N}_u = N+n_u$ and $\mathcal{N}_d = N+n_d$, respectively, are
not necessarily equal. In these models, the charged and neutral current terms of
the Lagrangian in the weak eigenstate basis can be written in matrix notation as
\begin{eqnarray}
{\mathcal L}_W & = & - \frac{g}{\sqrt 2}
\,\bar u_L^{(d)} \gamma^\mu d_L^{(d)} \,W_\mu^+ +\mathrm{h.c.} \,, \nonumber \\
{\mathcal L}_Z & = & - \frac{g}{2 c_W}
\left( \bar u_L^{(d)} \gamma^\mu u_L^{(d)} 
- \bar d_L^{(d)} \gamma^\mu d_L^{(d)}  
- 2 s_W^2 J_\mathrm{EM}^\mu \right) Z_\mu \,, \label{ec:1}
\end{eqnarray}
with $(u_L^{(d)},d_L^{(d)})$ doublets under $\mathrm{SU}(2)_L$ of dimension $N$
in flavour space. These terms have the same structure as in the SM, with $N$
generations of left-handed doublets in the isospin-related terms, but with all
the $\mathcal{N}_u$, $\mathcal{N}_d$ fields in $J_\mathrm{EM}$. The differences
show up in the mass eigenstate basis, where the Lagrangian reads
\begin{eqnarray}
{\mathcal L}_W & = & - \frac{g}{\sqrt 2}
\,\bar u_L \gamma^\mu V d_L \,W_\mu^+ +\mathrm{h.c.} \,, \nonumber \\
{\mathcal L}_Z & = & - \frac{g}{2 c_W} \left(
\bar u_L \gamma^\mu X^u u_L - \bar d_L \gamma^\mu X^d d_L
  - 2 s_W^2 J_\mathrm{EM}^\mu \right) Z_\mu \,.
\label{ec:2}
\end{eqnarray}
Here $u=(u,c,t,T,\dots)$ and $d=(d,s,b,B,\dots)$ are $\mathcal{N}_u$ and
${\mathcal N}_d$ dimensional vectors, and $X^u$, $X^d$ are matrices of dimension
$\mathcal{N}_u \times \mathcal{N}_u$, $\mathcal{N}_d \times \mathcal{N}_d$,
respectively. In general the $\mathcal{N}_u \times \mathcal{N}_d$ CKM matrix $V$
is neither unitary nor square.

The most distinctive feature of this class of models is the appearance of
tree-level FCN couplings in the mass eigenstate basis, originated by the mixing
of weak eigenstates with the same chirality and different isospin. These FCN
interactions mix left-handed fields, and are determined by the off-diagonal
entries in the matrices $X^u$, $X^d$. On the other hand, the diagonal $Zqq$
terms of up- or down-type mass eigenstates $q$ are (dropping here the
superscript on the $X$ matrices)
\begin{eqnarray}
c_L^q & = & \pm X_{qq} -2 Q_q s_W^2 \,, \nonumber \\
c_R^q & = & -2 Q_q s_W^2 \,,
\end{eqnarray}
with the plus (minus) sign for up (down) quarks.
With these definitions, the flavour-diagonal $Zqq$ vertices read
\begin{eqnarray}
{\mathcal L}_{Zqq} & = & - \frac{g}{2 c_W} \left(
\bar q_{} \gamma^\mu c_L^{q} q_{L} 
+ \bar q_{R} \gamma^\mu c_R^{q} q_{R} \right) \,.
\label{ec:2b}
\end{eqnarray}
For a SM-like mass eigenstate
without any left-handed singlet component, $X_{qq}=1$, $X_{qq'} = 0$ for
$q' \neq q$, and it has standard interactions with the $Z$ boson. For a mass
eigenstate with singlet components, $0 < X_{qq} < 1$, what implies nonzero FCN
couplings $X_{qq'}$ as well.

Let us write the unitary transformations between the mass and weak interaction
eigenstates,
\begin{eqnarray}
u_{L}^0 & = & \mathcal{U}^{uL} u_{L} \,, ~~~
u_{R}^0=\mathcal{U}^{uR} u_{R} \,, \nonumber \\
d_{L}^0 & = & \mathcal{U}^{dL} d_{L} \,, ~~~
d_{R}^0=\mathcal{U}^{dR} d_{R} \,,
\label{ec:3}
\end{eqnarray}
where $\mathcal{U}^{qL}$ and ${\mathcal U}^{qR}$ are $\mathcal{N}_q \times 
\mathcal{N}_q$ unitary matrices. The weak interaction eigenstates $q_{L,R}^0$
include doublets and singlets. It follows from Eqs.~(\ref{ec:1},\ref{ec:2}) that
\begin{eqnarray}
V_{\alpha \sigma} & = &
(\mathcal{U}^{uL}_{i\alpha})^*\, \mathcal{U}^{dL}_{i\sigma} \,,
  \nonumber \\[0.1cm]
X_{\alpha \beta}^{u} & = &
(\mathcal{U}^{uL}_{i\alpha})^*\, \mathcal{U}^{uL}_{i\beta} \,,~~~
X_{\sigma \tau}^d \; = \;
(\mathcal{U}^{dL}_{j\sigma})^*\, \mathcal{U}^{dL}_{j\tau} 
\label{ec:4}
\end{eqnarray}
with $i,j$ running over the left-handed doublets, $\alpha,\beta=u,c,t,T,\dots$
and $\sigma,\tau=d,s,b,B,\dots$. From these equations it is straightforward to
obtain the relations
\begin{eqnarray}
X^u & = & V \, V^\dagger \,, \nonumber \\
X^d & = & V^\dagger \, V \,,
\end{eqnarray}
and to observe that $X^u = (X^u)^\dagger$, $X^d = (X^d)^\dagger$. Furthermore,
we can see that in general $V$ is not an unitary matrix. We will restrict our
discussion to models where either $n_d=0$ or $n_u=0$, {\em i. e.} we will
consider either up singlets or down singlets, but not both at the same time. In
this context $V$ is a submatrix of a unitary matrix, and in each case we can
write 
\begin{eqnarray}
X^u_{\alpha \beta} & = & \sum_{i=1}^N V_{\alpha i} V_{\beta i}^* = 
\delta_{\alpha \beta} - \sum_{i=N+1}^{\mathcal{N}_u} V_{\alpha i} V_{\beta i}^*
\,, \nonumber \\
X^d_{\sigma \tau} & = & \sum_{i=1}^N V_{i \sigma}^* V_{i \tau} = 
\delta_{\sigma \tau} - \sum_{i=N+1}^{\mathcal{N}_d} V_{i \sigma}^*
   V_{i \tau} \,.
\label{ec:5}
\end{eqnarray}
It is enlightening to observe that for $\alpha \neq \beta$, $\sigma \neq \tau$,
we have the inequalities \cite{papiro16,papiro53}
\begin{eqnarray}
|X_{\alpha \beta}^u|^2 & \leq & (1-X_{\alpha \alpha}^u) (1-X_{\beta \beta}^u)
\,, \nonumber \\
|X_{\sigma \tau}^d|^2 & \leq & (1-X_{\sigma \sigma}^d) (1-X_{\tau \tau}^d) \,.
\label{ec:6}
\end{eqnarray}
Therefore, if for instance $X_{\alpha \alpha}^u=1$ (that is, if the diagonal
$Z$ vertex is the same as in the SM) the off-diagonal couplings involving the
quark $\alpha$ vanish. As a rule of thumb, FCN couplings arise at the expense of
decreasing the diagonal ones. This fact has strong implications on the limits on
FCN couplings, as we will later see.

The equality for $X^u$ in Eq.~(\ref{ec:6}) holds in particular if $n_u=1$.
Likewise, the equality for $X^d$ holds when $n_d=1$. This implies that the
introduction of {\em only one} extra singlet mixing significantly (as it is
usually done in the literature) results in additional restrictions in the
parameter space, and in principle may lead to different predictions. Moreover,
for $n_u=1$ or $n_d=1$ the CKM matrix has three independent CP violating phases,
whereas for $n_u=2$ or $n_d=2$ there are five. Hence, in our numerical analysis
we will consider also the situation when two singlets can have large mixing,
$n_u=2$, $n_d=0$ or $n_u=0$, $n_d=2$, to give a more complete picture. In the
rest of the paper we write the expressions for only one extra singlet for
simplicity.

\section{Direct limits}
\label{sec:3}
Even though in these SM extensions the $3 \times 3$ CKM matrix is not unitary,
in the two examples under study it is still a submatrix of a $4 \times 4$
unitary matrix $V$. The direct determination of the moduli of CKM matrix
elements \cite{papiro3} in Table~\ref{tab:0} not only sets direct limits on
these CKM elements themselves but also unitarity bounds on the rest. After the
requirement of $V_{tb} \sim 1$ from precision electroweak data (see
Section~\ref{sec:4}) these bounds are stronger. In this case, the Tevatron
constraint \cite{papiro4}
\begin{equation}
\frac{|V_{tb}|^2}{|V_{td}|^2+|V_{ts}|^2+|V_{tb}|^2} = 0.97^{+0.31}_{-0.24}
\label{ec:6b}
\end{equation}
is automatically satisfied.

\begin{table}[htb]
\begin{center}
\begin{tabular}{cc}
\hline
\hline
$|V_{ud}|$ & $0.9735 \pm 0.0008$ \\
$|V_{us}|$ & $0.2196 \pm 0.0023$ \\
$|V_{ub}|$ & $(3.6 \pm 1.0) \times 10^{-3}$ \\
$|V_{cd}|$ & $0.224 \pm 0.016$ \\
$|V_{cs}|$ & $0.97 \pm 0.11$ \\
$|V_{cb}|$ & $0.0402 \pm 0.0019$ \\
\hline
\hline
\end{tabular}
\caption{Direct measurements of CKM matrix elements. $V_{ub}$ is obtained from
$|V_{cb}|$ and the ratio $|V_{ub}/V_{cb}|$.
\label{tab:0}}
\end{center}
\end{table}

The non-observation of top decays $t \to qZ$, $q=u,c$ at Tevatron
\cite{papiro16b} provided the first limit on FCN couplings involving the top,
$|X_{qt}| \leq 0.84$ (from now on we omit the superscript when it is obvious).
These figures have improved with the analysis of single top production at LEP in
the process $e^+ e^- \to t \bar q + \bar t q$, which sets the bounds
$|X_{qt}| \leq 0.41$ \cite{papiro16c}. LEP limits are model-dependent because in
single top production there might possibly be contributions from a $\gamma tq$
effective coupling. These vertices are very small in most SM extensions, in
particular in models with quark singlets \cite{papiro16d}, thus in our case the
photon contribution may be safely ignored.

As long as new quarks have not been observed at Tevatron nor LEP, there are
various direct limits on their masses, depending on the decay channel analysed
\cite{papiro3}. We assume $m_T,m_B > 200$ GeV in our evaluations.

\section{Limits from precision electroweak data}
\label{sec:4}
\subsection{$R_b$, $R_c$ and FB asymmetries}
In the discussion after Eqs.~(\ref{ec:6}) we have observed that FCN interactions
can be bounded by examining the deviation from unity of the diagonal ones. This
is a particular example of a more general feature of these models, that the
isosinglet component of a mass eigenstate can be determined from its diagonal
couplings with the $Z$ boson. In this Section we will explain how the
experimental knowledge of $R_b$, $R_c$ and the FB asymmetries of the $b$ and
$c$ quarks constrains their mixing with isosinglets. We will study in detail the
case of the bottom quark; the discussion for the charm is rather alike.

$R_b$ is defined as the ratio
\begin{equation}
R_b = \frac{\Gamma(Z \to b \bar b)}
{\Gamma(Z \to \mathrm{hadrons})} \,.
\label{ec:7}
\end{equation}
The partial width to hadrons includes $u\bar u$, $d \bar d$, $s \bar s$,
$c \bar c$ and $b \bar b$. The numerator of this expression is proportional
to $(c_L^b)^2 + (c_R^b)^2$ plus a smaller term proportional to $m_b$. The pole
FB asymmetry of the $b$ quark is defined as
\begin{equation}
A_\mathrm{FB}^{0,b} = \frac{\sigma(\cos \theta > 0) - \sigma(\cos \theta < 0)}
{\sigma(\cos \theta > 0) + \sigma(\cos \theta < 0)} \,,
\label{ec:8}
\end{equation}
where $\theta$ is the angle between the bottom and the electron momenta in
the centre of mass frame
\footnote{These two observables {\em do not} include the photon contributions,
and $A_\mathrm{FB}^{0,b}$ is defined for massless external particles. They are
extracted from the experimental measurement of $e^+ e^- \to b \bar b$ after
correcting for the photon contribution, external masses and other effects
\cite{papiro1,papiro1b}.}.
The coupling parameter $\mathcal{A}_b$ of the bottom is defined as
\begin{equation}
\mathcal{A}_b = \frac{(c_L^b)^2 - (c_R^b)^2}{(c_L^b)^2 + (c_R^b)^2} \,.
\label{ec:8b}
\end{equation}
It is obtained from the left-right-forward-backward asymmetry of the $b$ quark
at SLD, and considered as an independent parameter in the fits, despite the fact
that the FB asymmetry can be expressed as
$A_\mathrm{FB}^{0,b}=3/4 \, \mathcal{A}_e \, \mathcal{A}_b$, with
$\mathcal{A}_e$ the coupling parameter of the electron.

At tree-level, $c_L^b = -X_{bb} + 2/3 s_W^2$, $c_R^q = 2/3 s_W^2$, hence in a
first approximation the mixing of the $b$ quark with down singlets in Model
$\modii$ decreases $X_{bb}$ from unity and thus decreases $R_b$,
$\mathcal{A}_b$ and $A_\mathrm{FB}^{0,b}$. The effect of some electroweak
radiative corrections can be taken into account using an
$\overline \mathrm{MS}$ definition of the sine of the weak angle,
$s_Z^2 = 0.23113$ \cite{papiro3} and  for the electron coupling an
``effective'' leptonic $\sin^2 \theta\,_\mathrm{lept}^\mathrm{eff} = 0.23137$
\cite{papiro1}. Other electroweak and QCD corrections that cannot be absorbed
into these definitions are included as well \cite{papiro17,papiro18}.
They are of order $0.6$\% for $u$, $c$ and $-0.25$ \% for $d$, $s$, $b$.
Furthermore, for the bottom quark there is an important correction originated by
triangle diagrams involving the top \cite{papiro19}:
\begin{equation}
\delta c_L^b = 2 \left( \frac{\alpha}{2 \pi} \right) |V_{tb}|^2 \; F(x_t)
\label{ec:9}
\end{equation}
(note that we use a different normalisation with respect to
Ref.~\cite{papiro19}), with $x_t=(m_t/M_W)^2$ and
\begin{eqnarray}
F(x_t) & = & \frac{1}{8 s_W^2} \left[ x_t + 2.880 \log x_t - 6.716 
+ (8.368  \log x_t - 3.408)/x_t \right. \nonumber \\
& & \left. + (9.126  \log x_t + 2.260)/x_t^2 
+ (4.043  \log x_t + 7.410)/x_t^3 + \dots \right] \,.
\label{ec:10}
\end{eqnarray}
We have omitted the imaginary part of $F(x_t)$ since it does not contribute to
$\delta c_L^b$. This large correction $\sim (m_t/M_W)^2$ is a consequence of the
non-decoupling behaviour of the top quark, and CKM suppression makes it relevant
only for the bottom. It decreases the value of $R_b$ by $4\sigma$ and has the
indirect effect of increasing $R_c$ slightly. Its inclusion is then crucial to
compare the theoretical calculation with experiment. In Table~\ref{tab:1} we
collect our SM predictions for $R_b$, $R_c$, $A_\mathrm{FB}^{0,b}$,
$A_\mathrm{FB}^{0,c}$, $\mathcal{A}_b$ and $\mathcal{A}_c$ calculated using the
parameters in Appendix~\ref{sec:ap1}, together with the experimental values
found in Ref.~\cite{papiro1}. The masses used are $\overline \mathrm{MS}$
masses at the scale $M_Z$. The correlation matrix necessary for the fit is in
Table~\ref{tab:2}.

\begin{table}[htb]
\begin{center}
\begin{tabular}{cccc}
\hline
\hline
& SM & Experimental & Total \\[-0.1cm]
& prediction & measurement & error \\
\hline
$R_b$   & $0.21558$ & $0.21646$ & $0.00065$ \\
$R_c$   & $0.1722$ & $0.1719$ & $0.0031$ \\
$A_\mathrm{FB}^{0,b}$ & $0.1039$ & $0.0990$ & $0.0017$ \\
$A_\mathrm{FB}^{0,c}$ & $0.0744$ & $0.0685$ & $0.0034$ \\
$\mathcal{A}_b$   & $0.935$ & $0.922$ & $0.020$ \\
$\mathcal{A}_c$   & $0.669$ & $0.670$ & $0.026$ \\[0.1cm]
\hline
\hline
\end{tabular}
\end{center}
\caption{SM calculation of $R_b$, $R_c$, $A_\mathrm{FB}^{0,b}$,
$A_\mathrm{FB}^{0,c}$, $\mathcal{A}_b$, $\mathcal{A}_c$ and experimental values.
\label{tab:1}}
\end{table}

\begin{table}[htb]
\begin{center}
\begin{tabular}{ccccccc}
\hline
\hline
 & $R_b$ & $R_c$ & $A_\mathrm{FB}^{0,b}$ & $A_\mathrm{FB}^{0,c}$ 
 & $\mathcal{A}_b$ & $\mathcal{A}_c$ \\[0.1cm]
\hline
$R_b$ & $1.00$ & $-0.14$ & $-0.08$ & $0.01$ & $-0.08$ & $0.04$ \\
$R_c$ & $-0.14$ & $1.00$ & $0.04$ & $-0.01$ & $0.03$ & $-0.05$ \\
$A_\mathrm{FB}^{0,b}$ & $-0.08$ & $0.04$ & $1.00$ & $0.15$ & $0.02$ & $0.00$ \\
$A_\mathrm{FB}^{0,c}$ & $0.01$ & $-0.01$ & $0.15$ & $1.00$ & $0.00$ & $0.01$ \\
$\mathcal{A}_b$ & $-0.08$ & $0.03$ & $0.02$ & $0.00$ & $1.00$ & $0.13$ \\
$\mathcal{A}_c$ & $0.04$ & $-0.05$ & $0.00$ & $0.01$ & $0.13$ & $1.00$ \\[0.1cm]
\hline
\hline
\end{tabular}
\end{center}
\caption{Correlation matrix for the experimental measurements of $R_b$, $R_c$,
$A_\mathrm{FB}^{0,b}$, $A_\mathrm{FB}^{0,c}$, $\mathcal{A}_b$ and
$\mathcal{A}_c$.
\label{tab:2}}
\end{table}

The mixing of the $b$ quark with down isosinglets decreases $V_{tb}$, making
this negative correction smaller in modulus. This is however less important than
the effect of the deviation of $X_{bb}$ from unity. The net effect is that in
Model $\modii$ $X_{bb}$, and hence also $V_{tb}$, are tightly constrained by
$R_b$ to be very close to unity.

In Model I the mixing of the top with singlets modifies the $Ztt$ interactions,
and the expression for $\delta c_L^b$ in Eq.~(\ref{ec:9}) must be corrected
accordingly (see Ref.~\cite{papiro20} and also Ref.~\cite{papiro20b}). The
decrease in $X_{tt}$ can be taken into account with the substitution
$F \to F+F_2$, with
\begin{equation}
F_2(x_t) = \frac{1}{8 s_W^2} \frac{X_{tt}-1}{2} \; x_t \left( 2 -
\frac{4}{x_t-1} \log x_t \right) \,.
\label{ec:11}
\end{equation}
Moreover, there are additional triangle diagrams with the mass eigenstate $T$
replacing the top, or involving $t$ and $T$. The $T$ quark contribution is
added to Eq.~(\ref{ec:9}) as the top term but multiplied by $|V_{Tb}|^2$. The
$t-T$ contribution is given by $V_{tb}^* V_{Tb} \,F_3(x_t,x_T)$, with
$x_T=(m_T/M_W)^2$ and
\footnote{In obtaining Eq.~(\ref{ec:12}) from the results quoted in
Ref.~\cite{papiro19} we have assumed a CKM parameterisation with
$V_{tb}^* V_{Tb}$ real. This is our case with the parameterisations used in the
numerical analysis in Section~\ref{sec:8}.}
\begin{eqnarray}
F_3(x_t,x_T) & = & \frac{1}{2 s_W^2}  \frac{\mathrm{Re}\,X_{tT}}{2} \left[
- \frac{1}{x_T-x_t} \left( \frac{{x_T}^2}{x_T-1} \log x_T 
-\frac{x_t^2}{x_t-1} \log x_t \right) \right.
\nonumber \\[0.1cm]
& & \left.+ \frac{x_t x_T}{x_T-x_t}  \left( \frac{x_T}{x_T-1} \log x_T 
-\frac{x_t}{x_t-1} \log x_t \right) \right] \,.
\label{ec:12}
\end{eqnarray}

In Model I this radiative correction gives the leading effect on $R_b$ of the
mixing. However, the presence of the new quark may make up for the difference in
the top contribution. Should the new mass eigenstate be degenerate with the top,
$m_T = m_t$ and $x_T = x_t$, one can verify that
\begin{eqnarray}
|V_{tb}|^2 \, F(x_t) + |V_{Tb}|^2 \, F(x_T) & = & 
|V_{tb}|^2 \left[ F(x_t)+F_2(x_t) \right] 
+ |V_{Tb}|^2 \left[ F(x_T)+F_2(x_T) \right] \nonumber \\
& & + V_{tb}^* V_{Tb} \, F_3(x_t,x_T) \,,
\label{ec:13}
\end{eqnarray}
as intuitively might be expected. Since
$(|V_{tb}|^2 + |V_{Tb}|^2) = |V_{tb}|_\mathrm{SM}^2$, this means that for
degenerate $t$, $T$ the correction has the same value as in the SM (and in this
situation the terms with $F_2$ and $F_3$ cancel each other). For $m_T \sim m_t$,
$\delta c_L^b$ has a similar magnitude as in the SM and low values
$V_{tb} \sim 0.6$ are allowed. For heavier $T$, the size of this radiative
correction sets limits on the CKM angle $V_{Tb}$, and thus on $V_{tb}$.

The study of the charm mixing and the constraints on its couplings from $R_c$,
$\mathcal{A}_c$ and $A_\mathrm{FB}^{0,c}$ is completely analogous (interchanging
the r\^ole of up and down singlets). In principle, the presence of a new heavy
down quark $B$ induces a large $m_B^2$-dependent correction, but this is
suppressed by the CKM factor $|V_{cB}|^2$ and hence the analysis is simplified.
The pole FB asymmetry of the quark $s$ has also been measured recently,
$A_\mathrm{FB}^{0,s} = 0.1008 \pm 0.0120$ \cite{papiro22}, though not nearly
with the same precision as the $b$ and $c$ asymmetries. This determination
assumes that the FB asymmetries of the $u,d$ quarks  and the $Z$ branching
ratios are fixed at their SM values and thus cannot be properly taken as a
direct measurement. We do not include it as a constraint, but anyway we have
checked that at this level of experimental precision it would not provide any
additional constraint on the model.

\subsection{Oblique parameters}
The oblique parameters $S$, $T$ and $U$ \cite{papiro21a,papiro21b} are used to
summarise the effects of new particles in weak currents in a compact form.
Provided these particles are heavy and couple weakly to the known fermions,
their leading effects in processes with only SM external particles are radiative
corrections given by vacuum polarization diagrams (oblique corrections), rather
than triangle and box diagrams (direct corrections). We will use the definitions
\cite{papiro23,papiro24}
\begin{eqnarray}
S & = & -16 \pi \frac{\Pi_{3Y}(M_Z^2)-\Pi_{3Y}(0)}{M_Z^2} \,, \nonumber \\
T & = & \frac{4 \pi}{M_Z^2 s_W^2 c_W^2} \left[\Pi_{11}(0)-\Pi_{33}(0) \right]
\,, \nonumber \\[0.2cm]
U & = & 16 \pi \left( \frac{\Pi_{11}(M_W^2)-\Pi_{11}(0)}{M_W^2}
 -\frac{\Pi_{33}(M_Z^2)-\Pi_{33}(0)}{M_Z^2} \right) \,.
\label{ec:15}
\end{eqnarray}
They are equivalent to the ones used in Ref.~\cite{papiro3}, as can be seen by
a change of basis. In these expressions only the contributions of new particles
are meant to be included. Radiative corrections from SM particles must be
treated separately because their leading effects are direct, not oblique. The
parameters $S$, $T$, $U$ are extracted from precision electroweak observables,
and their most recent values are in Table~\ref{tab:3}. The contributions to $S$,
$T$ and $U$ of an arbitrary number of families plus vector-like singlets and
doublets have been computed in Ref.~\cite{papiro24}. In our models there are no
exotic vector-like doublets, hence right-handed currents are absent and their
expressions simplify to
\begin{footnotesize}
\begin{eqnarray}
S & = & \frac{N_c}{2\pi} \left( \sum_{\alpha=1}^{\mathcal{N}_u}
\sum_{\sigma=1}^{\mathcal{N}_d} |V_{\alpha \sigma}|^2 \,
\psi_+(y_\alpha,y_\sigma)
- \sum_{\beta < \alpha}^{\mathcal{N}_u} |X^u_{\alpha \beta}|^2 \,
\psi_+(y_\alpha,y_\beta)
- \sum_{\tau < \sigma}^{\mathcal{N}_d} |X^d_{\sigma \tau}|^2 \,
\psi_+(y_\sigma,y_\tau) \right) \,, \nonumber \\
T & = & \frac{N_c}{16\pi s_W^2 c_W^2} \left( \sum_{\alpha=1}^{\mathcal{N}_u}
\sum_{\sigma=1}^{\mathcal{N}_d} |V_{\alpha \sigma}|^2 \,
\theta_+(y_\alpha,y_\sigma)
- \sum_{\beta < \alpha}^{\mathcal{N}_u} |X^u_{\alpha \beta}|^2 \,
\theta_+(y_\alpha,y_\beta)
- \sum_{\tau < \sigma}^{\mathcal{N}_d} |X^d_{\sigma \tau}|^2 \,
\theta_+(y_\sigma,y_\tau) \right) \,, \nonumber \\
U & = & -\frac{N_c}{2\pi} \left( \sum_{\alpha=1}^{\mathcal{N}_u}
\sum_{\sigma=1}^{\mathcal{N}_d} |V_{\alpha \sigma}|^2 \,
\chi_+(y_\alpha,y_\sigma)
- \sum_{\beta < \alpha}^{\mathcal{N}_u} |X^u_{\alpha \beta}|^2 \,
\chi_+(y_\alpha,y_\beta)
- \sum_{\tau < \sigma}^{\mathcal{N}_d} |X^d_{\sigma \tau}|^2 \,
\chi_+(y_\sigma,y_\tau) \right) \,,
\label{ec:17}
\end{eqnarray}
\end{footnotesize}\noindent
where $N_c=3$ is the number of colours, $y_i = (m_i/M_Z)^2$ and we use the
$\overline \mathrm{MS}$ definition of $s_W^2$, as well as
$\overline \mathrm{MS}$ masses at the scale $M_Z$. The functions multiplying the
mixing angles are
\begin{eqnarray}
\psi_+(y_1,y_2) & = & \frac{22 y_1+14 y_2}{9} - \frac{1}{9} \log \frac{y_1}{y_2}
+ \frac{11 y_1+1}{18} f(y_1,y_1) + \frac{7 y_2-1}{18} f(y_2,y_2) \,,
\nonumber \\
\theta_+(y_1,y_2) & = & y_1 + y_2 -\frac{2 y_1 y_2}{y_1-y_2} \log
\frac{y_1}{y_2} \,, \nonumber \\
\chi_+(y_1,y_2) & = & \frac{y_1+y_2}{2} - \frac{(y_1-y_2)^2}{3} +
\left[ \frac{(y_1-y_2)^3}{6} -\frac{y_1^2+y_2^2}{2(y_1-y_2)} \right] \log
\frac{y_1}{y_2} \nonumber \\
& & + \frac{y_1-1}{6} f(y_1,y_1) + \frac{y_2-1}{6} f(y_2,y_2) \nonumber \\
& & + \left[ \frac{1}{3} - \frac{y_1+y_2}{6} - \frac{(y_1-y_2)^2}{6} \right]
f(y_1,y_2) \,.
\label{ec:18}
\end{eqnarray}
The function $f$ is defined as
\begin{equation}
f(y_1,y_2) = \left \{ \mbox{\begin{tabular}{lcc}
$-2 \sqrt \Delta \left( \arctan \frac{y_1-y_2+1}{\sqrt \Delta}
- \arctan \frac{y_1-y_2-1}{\sqrt \Delta} \right)$
 & ~~ & $\Delta > 0$ \\[0.6cm]
$\sqrt{-\Delta} \log \frac{y_1+y_2-1+\sqrt{-\Delta}}{y_1+y_2-1-\sqrt{-\Delta}}$
 & ~~ & $\Delta \leq 0$
\end{tabular}} \right. \,,
\label{ec:19}
\end{equation}
with $\Delta = -1 - y_1^2 - y_2^2 + 2 y_1 + 2 y_2 + 2 y_1 y_2$. The functions
$\psi$, $\theta$, $\chi$ are symmetric under the interchange of their variables,
and $\theta$, $\chi$ satisfy $\theta(y,y)=0$, $\chi(y,y)=0$.

\begin{table}[htb]
\begin{center}
\begin{tabular}{cr}
\hline
\hline
$S$ & $-0.03 \pm 0.11$ \\
$T$ & $-0.02 \pm 0.13$ \\
$U$ & $0.24 \pm 0.13$ \\
\hline
\hline
\end{tabular}
\caption{Experimental values of the oblique parameters.
\label{tab:3}}
\end{center}
\end{table}

These expressions are far from transparent, and to have a better understanding
of them we will examine the example of an up singlet mixing exclusively with the
top. In this limit, the new contributions are
\begin{eqnarray}
S & = & \frac{N_c}{2\pi} \left\{ |V_{Tb}|^2 \left[ \psi_+(y_T,y_b) -
\psi_+(y_t,y_b) \right] - |X_{Tt}|^2 \psi_+(y_T,y_t) \right\} \,, \nonumber \\
T & = & \frac{N_c}{16\pi s_W^2 c_W^2} \left\{ |V_{Tb}|^2 \left[ 
\theta_+(y_T,y_b) - \theta_+(y_t,y_b) \right] 
- |X_{Tt}|^2 \theta_+(y_T,y_t) \right\} \,, \nonumber \\[0.1cm]
U & = & -\frac{N_c}{2\pi} \left\{ |V_{Tb}|^2 \left[ \chi_+(y_T,y_b) -
\chi_+(y_t,y_b) \right] - |X_{Tt}|^2 \chi_+(y_T,y_t) \right\} \,.
\label{ec:20}
\end{eqnarray}
The factors $|V_{Tb}|^2$, $|X_{Tt}|^2$ describing the mixing of the quark $T$
are not independent: as can be seen from the results in Section~\ref{sec:2},
for $n_u=1$ we have the relation
\begin{equation}
|X_{Tt}|^2 = |V_{Tb} V_{tb}|^2 = |V_{Tb}|^2 \, (1-|V_{Tb}|^2) \,.
\end{equation}
For $t$ and $T$ degenerate, $T$ and $U$ would automatically vanish independently
of the mixing, and $S=-0.16 \, |X_{Tt}|^2$. In order to obtain a simple
approximate formula when $m_T \gg m_t$, we approximate $1-|V_{Tb}|^2 \sim 1$
and keep only the leading order in $y_T$. (Needless to say, we use
Eqs.~(\ref{ec:17}--\ref{ec:19}) for our fits.) Using the numerical values of
$y_b$, $y_t$, this yields
\begin{eqnarray}
S & = & \frac{N_c}{2\pi} |V_{Tb}|^2 \left[ -0.34 + O(y_T^{-1}) \right] \,,
\nonumber \\
T & = & \frac{N_c}{16\pi s_W^2 c_W^2} |V_{Tb}|^2 \left[ -18.4 + 7.8 \log y_T
+ O(y_T^{-1}) \right] \,, \nonumber \\
U & = & -\frac{N_c}{2\pi} |V_{Tb}|^2 \left[ -0.60 + O(y_T^{-1}) \right] \,.
\label{ec:21}
\end{eqnarray}
These expressions give a fair estimate of the effect of the top mixing in the
oblique parameters. We notice that the effect on $S$, $U$ is very small,
$S = -0.16 \, |V_{Tb}|^2$, $U = 0.29 \, |V_{Tb}|^2$, but sizeable for $T$ (for
instance, $T = 2.7 \, |V_{Tb}|^2$ for $\overline{m}_T=500$ GeV). Indeed, the $T$
parameter bounds the CKM matrix element $V_{Tb}$ (and hence $V_{tb}$) more
effectively than the radiative correction to $R_b$ and better than low energy
observables.

In Model $\modii$ the analysis is similar, but the constraints from $R_b$ and
FCN processes at low energies are much more restrictive than these from oblique
corrections.

\section{Limits from FCN processes at low energies}
\label{sec:5}
In this Section we discuss low energy processes involving meson mixing and/or
decays. An important point is that almost all the observables analysed receive
short-distance contributions from box and/or penguin diagrams with $Q=2/3$ quark
loops (otherwise it will be indicated explicitly). The top amplitudes are
specially relevant due to the large top mass, and are proportional to
$V_{td}^* V_{ts}$, $V_{td}^* V_{tb}$ or $V_{ts}^* V_{tb}$ (or their squares),
depending on the meson considered. The observables are then sensitive to
$V_{td}$ and $V_{ts}$. (Also to $V_{tb}$, but the most important restrictions on
its modulus come from precision measurements examined in last Section.)
Additionally, there are extra contributions in the models under study: either
new box and penguin diagrams with an internal $T$ quark in Model I or diagrams
with tree-level  flavour-changing neutral currents (FCNC) mediated by the $Z$
boson in Model $\modii$. In any case, the new terms depend on products of two
elements of the fourth row of $V$ ($V_{Td}^* V_{Ts}$, $V_{Td}^* V_{Tb}$ or
$V_{Ts}^* V_{Tb}$ in Model I and FCN couplings $X_{ds}$, $X_{db}$ or $X_{sb}$ in
Model $\modii$).

The {\em a priori} unknown top and new physics terms are added coherently in
the expressions of all these observables. Then, in principle there may exist
a ``conspiracy'' between top and new physics contributions, with the first very
different from the SM prediction and new physics making up for the difference.
As long as we use a sufficiently exhaustive set of low energy observables and
reproduce their experimental values, this possibility can be limited. This is
because the products $V_{td}^* V_{ts}$, \dots, $V_{Td}^* V_{Ts}$ or $X_{ds}$,
etc. appear in the expressions of these observables in combinations with
different coefficients.

Our observables for Models I and $\modii$ include $|\delta m_B|$,
$|\delta m_{B_s}|$, $\varepsilon$, $\varepsilon'/\varepsilon$, the branching
ratios for $b \to s e^+ e^-$, $b \to s \mu^+ \mu^-$,
$K^+ \to \pi^+ \nu \bar \nu$, $K_L \to \mu^+ \mu^-$ and the CP asymmetry
$a_{\psi K_S}$. For model I we use $|\delta m_D|$ as well. It must be stressed
that {\em they are all independent} and give additional information that cannot
be obtained from the rest. For example, if we remove $\varepsilon$ from the list
we can find choices of parameters of our models for which all the remaining
observables agree with experiment (the precise criteria of agreement used will
be specified in Section~\ref{sec:8}) but $\varepsilon$ is more than 5 standard
deviations from its measured value. This procedure applied to each one shows
that none of them can be dismissed.

Once the values of the observables in these sets are in agreement with
experiment, the predictions for the mass difference $|\delta m_K|$ and some
other partial rates, like $b \to s \gamma$, $B \to s \nu \bar \nu$,
$B \to \mu^+ \mu^-$, $B_s \to \mu^+ \mu^-$, agree with SM expectations
($b \to s \gamma$ is nevertheless included in the fits). An important exception
is the decay $K_L \to \pi^0 \nu \bar \nu$, which will be studied in
Section~\ref{sec:7}. Several CP asymmetries can also differ from SM
expectations, and are thus good places to search for departures from the SM or
further restrict the models under consideration.

In the rest of this Section we review the theoretical calculation within Models
I and $\modii$ of the observables listed above, together with their experimental
status.

\subsection{Neutral meson oscillations}
\subsubsection{The $\Delta F=2$ effective Lagrangians}
The complete Lagrangian for $Q=-1/3$ external quarks in the presence of extra
down singlets has been obtained in Ref.~\cite{papiro25} and we follow their
discussion except for small changes in the notation. We ignore QCD corrections
for the moment and neglect external masses. The Lagrangians for $K^0$, $B^0$ and
$B_s^0$ oscillations are similar up to CKM factors, and for simplicity in the
notation we refer to the kaon system. The box contributions can be written as
\begin{eqnarray}
\mathcal{L}_\mathrm{eff}^\mathrm{box} & = & - \frac{G_F}{\sqrt 2}
\frac{ \alpha}{4 \pi s_W^2} \left[  \sum_{\alpha,\beta=u,c,t}
\lambda_{sd}^\alpha \lambda_{sd}^\beta \,
F(x_\alpha,x_\beta) \right]
\left( \bar s_L \gamma^\mu d_L \right) \,
\left( \bar s_L \gamma_\mu d_L \right) \,,
\label{ec:22}
\end{eqnarray}
with $\lambda_{sd}^\alpha = V_{\alpha s}^* V_{\alpha d}$, etc. The function
$F$ is not gauge-invariant, and its expression in the 't Hooft-Feynman gauge can
be found {\em e. g.} in Ref.~\cite{papiro26}. (This and other Inami-Lim
\cite{papiro27} functions are collected in Appendix~\ref{sec:ap2}.) The terms
involving the $u$ quark can be eliminated using
\begin{equation}
\sum_\alpha \lambda_{sd}^\alpha = X_{sd}
\end{equation}
and setting $x_u=0$, resulting in
\begin{eqnarray}
\mathcal{L}_\mathrm{eff}^\mathrm{box} & = & - \frac{G_F}{\sqrt 2}
\frac{ \alpha}{4 \pi s_W^2} \left[  \sum_{\alpha,\beta=c,t}
\lambda_{sd}^\alpha \lambda_{sd}^\beta \,
S_0(x_\alpha,x_\beta)
+ 8\, X_{sd} \sum_{\alpha=c,t} \lambda_{sd}^\alpha B_0(x_\alpha) 
+ X_{sd}^2 \right] \nonumber \\[0.2cm]
& & \times \left( \bar s_L \gamma^\mu d_L \right) \,
\left( \bar s_L \gamma_\mu d_L \right) \,,
\label{ec:23}
\end{eqnarray}
where the gauge-independent function $S_0$ is given in terms of the true box
function $F$ by
\begin{eqnarray}
S_0(x_\alpha,x_\beta) & = & F(x_\alpha,x_\beta)-F(x_\alpha,0)-
F(0,x_\beta) + F(0,0) \,,
\label{ec:24}
\end{eqnarray}
and $B_0$ is given in terms of $F$ by
\begin{equation}
4\, B_0(x_\alpha) = F(x_\alpha,0)-F(0,0) \,.
\label{ec:25}
\end{equation}
In addition there are two terms to be included in the Lagrangian. The first
corresponds to $Z$ tree-level FCNC,
\begin{equation}
\mathcal{L}_\mathrm{eff}^Z = - \frac{G_F}{\sqrt 2} X_{sd}^2
\left( \bar s_L \gamma^\mu d_L \right) \,
\left( \bar s_L \gamma_\mu d_L \right) \,.
\label{ec:26}
\end{equation}
The second is originated from diagrams with one tree-level FCN coupling and one
triangle loop. Its contribution plus the $B_0$ term in Eq.~(\ref{ec:23}) can be
compared with the short-distance effective Lagrangian for $K^0 \to \mu^+ \mu^-$
(see Ref.~\cite{papiro25} for the details), concluding that the sum of both
gives the gauge-invariant Inami-Lim function $Y_0$ with a minus sign. The full
gauge-invariant $\Delta S=2$ effective Lagrangian then reads
\begin{eqnarray}
\mathcal{L}_\mathrm{eff}^\modii & = & - \frac{G_F}{\sqrt 2} \left[ 
\frac{ \alpha}{4 \pi s_W^2} \sum_{\alpha,\beta=c,t}
\lambda_{sd}^\alpha \lambda_{sd}^\beta \, S_0(x_\alpha,x_\beta)
- 8\, \frac{ \alpha}{4 \pi s_W^2} X_{sd}
  \sum_{\alpha=c,t} \lambda_{sd}^\alpha Y_0(x_\alpha) + X_{sd}^2  \right]
\nonumber \\[0.2cm]
& & \times \left( \bar s_L \gamma^\mu d_L \right) \,
\left( \bar s_L \gamma_\mu d_L \right) \,.
\label{ec:27}
\end{eqnarray}
We use the $\modii$ superscript to refer to model $\modii$. The $X_{ds}^2$ term
in Eq.~(\ref{ec:22}) is subleading with respect to $\mathcal{L}_\mathrm{eff}^Z$,
and it has been omitted.

The SM Lagrangian can be readily recovered setting $X_{ds}=0$ in the above
equation. In Model I the Lagrangian reduces to the SM-like box contributions but
with terms involving the new mass eigenstate $T$,
\begin{eqnarray}
\mathcal{L}_\mathrm{eff}^\mathrm{I} & = & - \frac{G_F}{\sqrt 2} \left[ 
\frac{ \alpha}{4 \pi s_W^2} \sum_{\alpha,\beta=c,t,T}
  \lambda_{sd}^\alpha \lambda_{sd}^\beta \, S_0(x_\alpha,x_\beta)
 \right] \left( \bar s_L \gamma^\mu d_L \right) \,
\left( \bar s_L \gamma_\mu d_L \right) \,.
\label{ec:28}
\end{eqnarray}
In the $B^0$ and $B_s^0$ systems the approximation of vanishing external masses
is not justified for the $b$ quark. However, the two terms involving
$S_0(x_c) \equiv S_0(x_c,x_c)$ and $S_0(x_c,x_t)$ are much smaller than the one
with $S_0(x_t) \equiv S_0(x_t,x_t)$ and can be neglected, and for the latter
$m_b \ll m_t$ and the approximation is valid. The effective Lagrangian for
$D^0 - \bar D^0$ mixing is more problematic and we will deal with it later.

Short-distance QCD corrections are included in these Lagrangians as $\eta$
factors multiplying each term in the usual way. These factors account for high
energy QCD effects and renormalisation group (RG) evolution to lower scales
\cite{papiro29}. When available, we use next-to-leading order (NLO) corrections
\cite{papiro30}. For the nonstandard contributions we use leading logarithmic
(LL) RG evolution \cite{papiro31}. The differences between LL and NLO
corrections are minimal provided we use $\overline\mathrm{MS}$ masses
$\overline{m}_i(m_i)$ in the evaluations \cite{papiro30}. Some representative
QCD corrections for the new terms are
\begin{eqnarray}
\eta^K_Z & = & \left[ \alpha_s(m_c) \right]^\frac{6}{27}
\left[ \frac{\alpha_s(m_b)}{\alpha_s(m_c)} \right]^\frac{6}{25}
\left[ \frac{\alpha_s(M_Z)}{\alpha_s(m_b)} \right]^\frac{6}{23} \,,
\nonumber \\ 
\eta^B_{TT} & = & \left[ \alpha_s(m_t) \right]^\frac{6}{23}
\left[ \frac{\alpha_s(m_T)}{\alpha_s(m_t)} \right]^\frac{6}{21} \,.
\label{ec:31}
\end{eqnarray}
Here the superscripts $K$, $B$ refer to the neutral mesons, and the subscripts
to the term considered.

\subsubsection{$K^0$ oscillations}
The element $M_{12}$ of the $K^0 - \bar K^0$ mixing matrix is obtained from the
effective Lagrangian (see for instance Ref.~\cite{papiro29}),
\begin{eqnarray}
M_{12}^K & = & \frac{G_F^2 M_W^2 f_K^2 {\hat B}_K m_{K^0}}{12 \pi^2} \left[
(\lambda_{ds}^c)^2 \eta^K_{cc} S_0(x_c)
+ (\lambda_{ds}^t)^2 \eta^K_{tt} S_0(x_t)
\right. \nonumber \\[0.2cm]
& & \left. + 2 \lambda_{ds}^c \lambda_{ds}^t \eta^K_{ct} S_0(x_c,x_t)
+ \Delta_K \right] \,.
\label{ec:32}
\end{eqnarray}
In this expression $m_{K^0} = 498$ MeV is the $K^0$ mass, $f_K = 160$ MeV the
kaon decay constant taken from experiment and  ${\hat B}_K = 0.86 \pm 0.15$
\cite{papiro32} is the bag parameter. The QCD corrections are
$\eta_{cc}^K= 1.38 \pm 0.20$, $\eta_{tt}^K = 0.57$, $\eta_{ct}^K = 0.47$
\cite{papiro33} (we do not explicitly write the errors when they are
negligible). The extra piece $\Delta_K$ in models I and $\modii$ is
\begin{eqnarray}
\Delta_K^\mathrm{I} & = & (\lambda_{ds}^T)^2 \eta^K_{TT} S_0(x_T) +
2 \lambda_{ds}^c \lambda_{ds}^T \eta^K_{cT} S_0(x_c,x_T) 
+ 2 \lambda_{ds}^t \lambda_{ds}^T \eta^K_{tT} S_0(x_t,x_T)
\nonumber \,, \\[0.2cm]
\Delta_K^\modii & = & -8 X_{ds} \left[ \lambda_{ds}^c \eta_Z^K Y_0(x_c)
+ \lambda_{ds}^t \eta_{tt}^K Y_0(x_t) \right] +\frac{4 \pi s_W^2}{\alpha} 
 \eta_Z^K X_{ds}^2 \,,
\label{ec:33}
\end{eqnarray}
with $\eta_{TT}^K = 0.58$, $\eta_Z^K = 0.60$. We estimate
$\eta_{cT}^K \simeq \eta_{ct}^K$, $\eta_{tT}^K \simeq \eta_{TT}^K$, and expect
that this is a good approximation because RG evolution is slower at larger
scales. 

In the neutral kaon system the mass difference $\delta m_K$ can be written as
\begin{equation}
\delta m_K = 2\, \mathrm{Re}\, M_{12}^K + \delta m_K^\mathrm{LD}
\label{ec:34}
\end{equation}
where the second term is a long-distance contribution that cannot be calculated
reliably. For the first term we obtain
$(4.64 \pm 0.68) \times 10^{-3}$ ps$^{-1}$ within the SM,
whereas $\delta m_K = 5.30 \times 10^{-3}$ ps$^{-1}$. The large
$\sim 30$ \%
long-distance contribution prevents us from using $\delta m_K$ as a constraint
on our models, but we observe anyway that the short-distance part
$2\, \mathrm{Re}\, M_{12}^K$ always takes values very close to the SM prediction
once all other constraints are fulfilled.

The CP violating parameter $\varepsilon$ is calculated as
\footnote{This expression assumes a phase convention where $V_{ud}^* V_{us}$
is real. For a rephasing-invariant definition of $\varepsilon$ see
Ref.~\cite{libro}.}
\begin{equation}
\varepsilon = e^{i \pi/4} \frac{\mathrm{Im}\, M_{12}^K}{\sqrt 2 \;
  \delta m_K} \,
\label{ec:35}
\end{equation}
and in the SM it is close to its experimental value
$(2.282 \pm 0.017) \times 10^{-3}$ after a proper choice of the CKM phase
$\delta$ (see Appendix~\ref{sec:ap1}). The SM prediction with the phase
$\delta = 1.014$ that best fits $\varepsilon$, $\varepsilon'/\varepsilon$,
$a_{\psi K_S}$ and $|\delta m_B|$ is
$\varepsilon = (2.18 \pm 0.38) \times 10^{-3}$. Notice that there is a large
theoretical error in the calculation, mainly a consequence of the uncertainty in
$\hat B_K$, which results in a poor knowledge of the CKM phase that reproduces
$\varepsilon$ within the SM. In Models I and $\modii$ this parameter receives
contributions from several CP violating phases and thus it cannot be used to
extract one in particular. Instead, we let the phases arbitrary and require that
the prediction for $\varepsilon$ agrees with experiment.

\subsubsection{$B^0$ oscillations}
The element $M_{12}^B$ of the $B^0 - \bar B^0$ mixing matrix is
\begin{eqnarray}
M_{12}^B & = & \frac{G_F^2 M_W^2 f_B^2 {\hat B}_B m_{B^0}}{12 \pi^2} \left[
(\lambda_{db}^t)^2 \eta^B_{tt} S_0(x_t) + \Delta_B \right]
\label{ec:36}
\end{eqnarray}
with $m_{B^0} = 5.279$ GeV. We use $f_B = 200 \pm 30$ MeV,
${\hat B}_B = 1.30 \pm 0.18$ from lattice calculations \cite{papiro34}.
The terms corresponding to $S_0(x_c)$ and $S_0(x_c,x_t)$ have been discarded
as usual, because in the SM, as well as in our models, they are numerically
$2-3$ orders of magnitude smaller than the $S_0(x_t)$ term (the CKM angles are
of the same order and the $S_0$ functions are much smaller). The QCD correction
is $\eta_{tt}^B = 0.55$. The nonstandard contributions are
\begin{eqnarray}
\Delta_B^\mathrm{I} & = & (\lambda_{db}^T)^2 \eta^B_{TT} S_0(x_T) +
2 \lambda_{db}^t \lambda_{db}^T \eta^B_{tT} S_0(x_t,x_T) \,,
\nonumber \\[0.1cm]
\Delta_B^\modii & = & -8 X_{db} \lambda_{db}^t \eta_{tt}^B Y_0(x_t)
+ \frac{4 \pi s_W^2}{\alpha} \eta_Z^B X_{db}^2 \,.
\label{ec:37}
\end{eqnarray}
The terms $S_0(x_c,x_T)$ in $\Delta_B^\mathrm{I}$ and $Y_0(x_c)$ in
$\Delta_B^\modii$ have been dropped with the same argument as above. The QCD
corrections for the rest are $\eta_{TT}^B = 0.55$, $\eta_Z^B = 0.57$ and we
approximate $\eta_{tT}^B \simeq \eta_{TT}^B$.

Since $|\Gamma_{12}^B| \ll |M_{12}^B|$, the mass difference in the $B$ system is
\begin{equation}
|\delta m_B| = 2 \, |M_{12}^B| \,,
\label{ec:38}
\end{equation}
and is useful to constrain $\lambda_{db}^t$ and the new physics parameters,
$\lambda_{db}^T$ or $X_{db}$ depending on the model considered. Long-distance
effects are negligible in the $B$ system, and the SM calculation yields 
$|\delta m_B| = 0.49 \pm 0.16$ ps$^{-1}$, to be compared with the experimental
value $0.489 \pm 0.008$ ps$^{-1}$.

A second restriction regarding $B$ oscillations comes from the time-dependent
asymmetry  in the decay $B^0 \to \psi K_S$ (see for instance Ref.~\cite{libro}
for a precise definition). This process is mediated by the quark-level
transition $\bar b \to \bar c c \bar s$ and takes place at tree-level, with
small penguin corrections. The amplitude for the decay can then be written to a
good approximation as $A  = \tilde A\, V_{cb}^* V_{cs}$, with $\tilde A$ real.
Therefore the asymmetry is \cite{papiro5}
$a_{\psi K_S} = \mathrm{Im}\,\lambda_{\psi K_S}$, with
\begin{eqnarray}
\lambda_{\psi K_S} & = & - \frac{(M_{12}^B)^*}{|M_{12}^B|} \,
\frac{\bar A}{A} \, \frac{M_{12}^K}{|M_{12}^K|}
 = - \frac{(M_{12}^B)^*}{|M_{12}^B|} \,
\frac{V_{cb} V_{cs}^*}{V_{cb}^* V_{cs}} \, \frac{M_{12}^K}{|M_{12}^K|}
\label{ec:39}
\end{eqnarray}
and provides a constraint on the combination of phases of $B$, $K$ mixing and
the decay $\bar b \to \bar c c \bar s$, which are functions of the CKM CP
violating phases and angles. Our calculation within the SM gives
$a_{\psi K_S} = 0.71$, and with other choices of parameters for the CKM matrix
the prediction may change in $\pm 0.08$. The world average is
$a_{\psi K_S}= 0.734 \pm 0.054$ \cite{papiro36}. This asymmetry can also be
expressed as
\begin{equation}
a_{\psi K_S} = \sin (2\beta+2\theta_B-2\theta_K) \,,
\label{ec:40}
\end{equation}
with $\beta$ one of the angles of the well-known $db$ unitarity triangle,
\begin{equation}
\beta = \arg \left[ -\frac{V_{cd} V_{cb}^*}{V_{td} V_{tb}^*} \right] \,,
\label{ec:41}
\end{equation}
and $\theta_B$, $\theta_K$ parameterising the deviation of the mixing amplitude
phases with respect to the SM,
\begin{equation}
2\,\theta_B = \arg \frac{M_{12}^B}{(M_{12}^B)_\mathrm{SM}} \,, ~~~
2\,\theta_K = \arg \frac{M_{12}^K}{(M_{12}^K)_\mathrm{SM}} \,.
\label{ec:42}
\end{equation}
In the absence of new physics, or if the extra phases $\theta_B$, $\theta_K$
cancel, $a_{\psi K_S} = \sin 2\beta$.

The phase of $M_{12}^K$ is relatively fixed by the determination of
$\varepsilon$ and $\delta m_K$. Despite the good experimental precision of both
measurements,  the former has a large theoretical uncertainty from $\hat B_K$
and the latter from long-distance contributions. This allows $\theta_K$ to be
different from zero, but it must be small anyway. The agreement of
$a_{\psi K_S}$ with experiment then constrains the phase $\theta_B$. The
asymmetry in semileptonic decays depends also on $\theta_B$ \cite{papiro36b} and
does not provide any extra constraint on the parameters of these models.

\subsubsection{$B^0_s$ oscillations}
The analysis of $B_s^0$ oscillations is very similar to the previous one for
the $B^0$ system,  with
\begin{eqnarray}
M_{12}^{B_s} & = & \frac{G_F^2 M_W^2 f_{B_s}^2 {\hat B}_{B_s} m_{B_s^0}}{12
\pi^2} \left[ (\lambda_{sb}^t)^2 \eta^{B_s}_{tt} S_0(x_t)
 + \Delta_{B_s} \right]
\end{eqnarray}
and $m_{B_s^0} = 5.370$ GeV, $f_{B_s} = 230 \pm 35$ MeV,
${\hat B}_{B_s} =  1.30 \pm 0.18$ \cite{papiro34}. The $S_0(x_c)$ and
$S_0(x_c,x_t)$ terms have again been neglected, and the QCD correction for the
$S_0(x_t)$ term is $\eta^{B_s}_{tt} = 0.55$. The new contributions are
\begin{eqnarray}
\Delta_{B_s}^\mathrm{I} & = & (\lambda_{sb}^T)^2 \eta^{B_s}_{TT} S_0(x_T) +
2 \lambda_{sb}^t \lambda_{sb}^T \eta^{B_s}_{tT} S_0(x_t,x_T) \,,
\nonumber \\[0.1cm]
\Delta_{B_s}^\modii & = & -8 X_{sb} \lambda_{sb}^t \eta_{tt}^{B_s} Y_0(x_t) 
+ \frac{4 \pi s_W^2}{\alpha} \eta_Z^{B_s} X_{sb}^2 \,.
\end{eqnarray}
with the terms involving $S_0(x_c,x_T)$ in $\Delta_B^\mathrm{I}$ and $Y_0(x_c)$
in $\Delta_{B_s}^\modii$ discarded. The QCD correction factors are the same as
for $M_{12}^B$. The SM estimate for $|\delta m_{B_s}| = 2 |M_{12}^{B_s}|$ is
$17.6 \pm 5.9$ ps$^{-1}$. Experimentally only a lower bound for
$|\delta m_{B_s}|$ exists, $|\delta m_{B_s}| \geq 13.1 $ ps$^{-1}$ with a 95\%
confidence level (CL), which can be saturated in Models I and $\modii$ and thus
provides a constraint not always considered in the literature. Larger values
than in the SM are also possible.

\subsubsection{$D^0$ oscillations}
In contrast with the $K^0$ and $B^0$ systems, $D^0$ mixing is mediated by box
diagrams with $Q=-1/3$ internal quarks. This circumstance leads to a very small
mass difference in the SM, as a consequence of the GIM mechanism. In addition,
the approximation of vanishing external masses is inconsistent, and with a
careful analysis including the charm mass an extra suppression
$\sim (m_s/m_c)^2$ is found \cite{papiro37,papiro38}, resulting in
$|\delta m_D| \sim 10^{-17}$ GeV. NLO contributions, for example dipenguin
diagrams \cite{papiro39}, are of the same order, but to our knowledge a full NLO
calculation is not available yet. Long-distance contributions are estimated to
be $|\delta m_D| \sim 10^{-16}$ GeV \cite{papiro40}.

On the other hand, the present experimental limit, $|\delta m_D| < 0.07$
ps$^{-1}$ $ = 4.6 \times 10^{-14}$ GeV with a 95 CL, is still orders of
magnitude above SM expectations. This limit can be saturated in Model I with a
tree-level FCN coupling $X_{cu}$ \cite{papiro41a}. In Model $\modii$ with a new
quark $B$ the GIM suppression is partially removed but we have checked that
$D^0$ mixing does not provide any additional constraint for a mass $m_B < 1$
TeV. Hence here we only discuss Model I. The element $M_{12}^D$ is then
\begin{equation}
M_{12}^D = \frac{G_F^2 M_W^2 f_D^2 {\hat B}_D m_{D^0}}{12 \pi^2} \left[
\frac{4 \pi s_W^2}{\alpha} \eta_Z^D X_{cu}^2 \right] \,,
\label{ec:43}
\end{equation}
where $m_{D^0} = 1.865$ GeV and $f_D = 215 \pm 15$ MeV \cite{papiro41}. We
assume ${\hat B}_D = 1.0 \pm 0.3$. We have omitted the SM terms, whose explicit
expression can be found for instance in Ref.~\cite{papiro38}, since they are
generically much smaller than the one written above.
In contrast with $K^0$ and
$B^0$ oscillations, the terms linear in $X_{cu}$ are both negligible due to the
small masses $m_s$, $m_b$, and we have dropped them
\footnote{Extending the discussion in Ref.~\cite{papiro25} to the case of
$D^0 - \bar D^0$ mixing, we can argue that the functions multiplying the terms
linear in $X_{cu}$ are the Inami-Lim functions appearing in
$D^0 \to \mu^+ \mu^-$, which are also in this case $-Y_0$ \cite{papiro74}.}.
The QCD correction is $\eta^D_Z = 0.59$. The mass difference is given by
$|\delta m_D| = 2 \, |M_{12}^D|$, and provides the most stringent limit on
$X_{cu}$.

\subsection{$K$ decays}
\subsubsection{$K^+ \to \pi^+ \nu \bar \nu$}
The importance of the rare kaon decay $K^+ \to \pi^+ \nu \bar \nu$ in setting
limits on the FCN coupling $X_{sd}$ has been pointed out before \cite{papiro53}.
This is a theoretically very clean process after NLO corrections reduce the
scale dependence. The uncertainty in the hadronic matrix element can be avoided
relating this process to the leading decay $K^+ \to \pi^0 e^+ \bar \nu$ using
isospin symmetry, and then using the measured rate for the latter:
\begin{eqnarray}
\frac{\mathrm{Br}(K^+ \to \pi^+ \nu \bar \nu)}
{\mathrm{Br}(K^+ \to \pi^0 e^+ \bar \nu)} & = & 
\frac{r_{K^+} \alpha^2}{2 \pi^2 s_W^4 |V_{us}|^2} 
\sum_{l=e,\mu,\tau} \left| \lambda_{sd}^c X_{NL}^l
+ \lambda_{sd}^t \eta_t^X X_0(x_t) + \Delta_{K^+} \right|^2 \,.
\label{ec:61}
\end{eqnarray}
The factor $r_{K^+} = 0.901$ \cite{papiro54} accounts for isospin breaking
corrections. The charm contributions at NLO are \cite{papiro55}
$X_{NL}^{e,\mu} = (10.6 \pm 1.5) \times 10^{-4}$,
$X_{NL}^{\tau} = (7.1 \pm 1.4) \times 10^{-4}$, and the QCD correction to the
top term is $\eta_t^X = 0.994$ \cite{papiro56}. The function $X_0 = C_0-4 B_0$
can be found in Appendix~\ref{sec:ap2}. The top and charm terms have similar
size because $X_0(x_t) \gg X_{NL}^l$ but
$\lambda_{sd}^t \ll \lambda_{sd}^c$. With
$\mathrm{Br}(K^+ \to \pi^0 e^+ \bar \nu) = 0.0487$ we obtain the SM value
$\mathrm{Br}(K^+ \to \pi^+ \nu \bar \nu) = (6.4 \pm 0.6) \times 10^{-11}$, where
in the uncertainty we only include that derived from $m_t$ and $X_{NL}^l$, and
not from CKM mixing angles. Experimentally there are only two
$K^+ \to \pi^+ \nu \bar \nu$ events \cite{papiro57}. The corresponding
90\% CL interval for the branching ratio is $[3.2,48] \times 10^{-11}$.

The new physics contributions are denoted by $\Delta_{K^+}$, and in Models I and
$\modii$ they read
\begin{eqnarray}
\Delta_{K^+}^\mathrm{I} & = & \lambda_{sd}^T \eta_T^X X_0(x_T) \,, \nonumber \\
\Delta_{K^+}^\modii & = & C_{U2Z} X_{sd} \,,
\label{ec:62}
\end{eqnarray}
where the factor $C_{U2Z}$ in $\Delta^\modii_{K^+}$ is \cite{papiro44}
\begin{equation}
C_{U2Z} = -\frac{\pi^2}{\sqrt 2 G_F M_W^2} = -\frac{\pi s_W^2}{\alpha} \,.
\label{ec:47}
\end{equation}
In Model I there is another consequence of the mixing of the top quark not
considered in these expressions: $X_{tt}$ and $X_{TT}$ are different from unity
(hence the function $C_0$ corresponding to $Z$ penguins changes) and there are
extra penguin diagrams with $T$ and $t$, proportional to the FCN coupling
$X_{tT}$. This is the same kind of modification that we have seen in the
discussion of the radiative correction to $R_b$. There it was found that the
net effect of the top mixing would cancel for $m_T = m_t$ and is small for
$m_T \sim m_t$. The magnitude of the correction required to take this effect
into account in $\mathrm{Br}(K^+ \to \pi^+ \nu \bar \nu)$ can be estimated in
analogy with that case. We find that the correction grows with $X_{tT}$ and
$m_T$; however, these cannot be both large, as required by oblique parameters.
The result is that the error made using Eqs.~(\ref{ec:61},\ref{ec:62}) is
smaller than the combined uncertainty from $X_{NL}$ and $m_t$ (10\%). For
$X_{tT}$ in its upper limit it amounts to a 6\% extra systematic error,
unimportant with present experimental precision. For each value of $X_{tT}$ and
$m_T$ we include the estimate of the correction required in the total
theoretical uncertainty. Bearing in mind the approximation done in using
Eqs.~(\ref{ec:61},\ref{ec:62}), we also omit the QCD factors in the calculation
because they represent a smaller effect.

This decay sets relevant constraints on $\lambda^t_{sd}$, $\lambda^T_{sd}$ and
$X_{sd}$ that cannot be obtained from the rest of processes studied in this
Section. This fact has been explicitly proved studying what the range of
predictions for  $\mathrm{Br}(K^+ \to \pi^+ \nu \bar \nu)$ would be if the rest
of the restrictions were fulfilled but not the one regarding
$K^+ \to \pi^+ \nu \bar \nu$. Since in some regions of the parameter space
$\mathrm{Br}(K^+ \to \pi^+ \nu \bar \nu)$ would be out of the
experimental interval, this process cannot be discarded in the analysis.

\subsubsection{$K_L \to \mu^+ \mu^-$}
A complementary limit on $\lambda^t_{sd}$, $\lambda^T_{sd}$ and $X_{sd}$ comes
from the short-distance contribution to the decay $K_L \to \mu^+ \mu^-$. 
Although theoretically this is a clean calculation, the extraction from actual
experimental data is very difficult. The branching ratio
$\mathrm{Br}(K_L \to \mu^+ \mu^-) = 7.18 \pm 0.17 \times 10^{-9}$
\cite{papiro58}, can be decomposed in a dispersive part $[\mathrm{Re} \, A]^2$
and an absorptive part $[\mathrm{Im} \, A]^2$. The imaginary part can be
calculated from $\mathrm{Br}(K_L \to \gamma \gamma)$ and amounts to
$(7.07 \pm 0.18) \times 10^{-9}$ \cite{papiro3}, which almost saturates the
total rate. The extraction of the long-distance component from the real part
$[\mathrm{Re} \, A]^2 = (1.1 \pm 2.4) \times 10^{-10}$ is not model-independent
\cite{papiro59}, but as long as our aim is to place limits on new physics we can
use the model in Ref.~\cite{papiro60} as an
estimate, obtaining the 90\% CL bound
$\mathrm{Re}\, A_\mathrm{SD}  \leq 1.9 \times 10^{-9}$.

On the theoretical side, the calculation of the short-distance part of the decay
is done relating it to $K^+ \to \mu^+ \nu$,
\begin{eqnarray}
\frac{\mathrm{Br}(K_L \to \mu^+ \mu^-)_\mathrm{SD}}
{\mathrm{Br}(K^+ \to \mu^+ \nu)} & = & \frac{\tau_{K_L}}{\tau_{K^+}}
\frac{\alpha^2}{\pi^2 s_W^4 |V_{us}|^2} \nonumber \\[0.1cm]
& & \times \left[ Y_{NL} \mathrm{Re}\,
\lambda_{sd}^c + \eta_t^Y Y_0(x_t) \, \mathrm{Re}\, \lambda_{sd}^t
+ \Delta_{K_L} \right]^2 \,,~
\label{ec:63}
\end{eqnarray}
with $\tau_{K_L} = 5.17 \times 10^{-8}$ s,
$\tau_{K^+} = 1.238 \times 10^{-8}$ s. The factor in the charm term is
$Y_{NL} = (2.94 \pm 0.28) \times 10^{-4}$ at NLO \cite{papiro55}. The function
$Y_0 = C_0-B_0$ can be found in Appendix~\ref{sec:ap2}, and the QCD correction
for the top is very close to unity, $\eta_t^Y = 1.012$ \cite{papiro56}. Using
$\mathrm{Br}(K^+ \to \mu^+ \nu) = 0.6343$ from experiment, the SM prediction
is $\mathrm{Br}(K_L \to \mu^+ \mu^-)_\mathrm{SD} = (6.6 \pm 0.6) \times
10^{-10}$. The new physics contributions are
\begin{eqnarray}
\Delta_{K_L}^\mathrm{I} & = & \eta_T^Y Y_0(x_T) \, \mathrm{Re}\, \lambda_{sd}^T
\,, \nonumber \\
\Delta_{K_L}^\modii & = & \mathrm{Re}\, C_{U2Z} X_{sd} \,.
\label{ec:64}
\end{eqnarray}
As in $K^+ \to \pi^+ \nu \bar \nu$ the mixing of the top quark modifies the
Inami-Lim function $C_0$ and adds a new $t-T$ term. The net effect is small and
has been taken into account in the theoretical uncertainty.

\subsection{B decays}
\subsubsection{$B \to X_s \gamma$}
The inclusive decay width $\Gamma (B \to X_s \gamma)$ can be well approximated
by the parton-level width $\Gamma (b \to s \gamma)$. In order to reduce
uncertainties, it is customary to calculate instead the ratio
\begin{equation}
R_\gamma \equiv \frac{\Gamma (b \to s \gamma)}{\Gamma (b \to c e \bar \nu)}
\label{ec:47b}
\end{equation}
and derive $\Gamma (b \to s \gamma)$ from $R_\gamma$ and the experimental
measurement of $\Gamma (b \to c e \bar \nu)$. The ratio $R_\gamma$ is given by
\begin{equation}
R_\gamma = \frac{|\lambda_{sb}^t|^2}{|V_{cb}|^2} \frac{6\alpha}{\pi f(z)}
|C_{7\gamma}(\mu)|^2 \,,
\label{ec:47c}
\end{equation}
where $z=m_c/m_b$ (pole masses) and
\begin{equation}
f(z) = 1 - 8 z^2 + 8 z^6 - z^8 -24 z^4 \log z
\label{ec:47d}
\end{equation}
is a phase space factor for $b \to c e \bar \nu$. The Wilson coefficient
$C_{7\gamma}(\mu)$ is obtained from the relevant coefficients at the scale $M_W$
by RG evolution.

The study of $b \to s \gamma$ in the context of SM extensions with up and down
quark singlets has been carried out at leading order (one loop) in
Refs.~\cite{papiro45,papiro46}. The Wilson coefficients at the scale $M_W$
relevant for this process are, using the notation of Ref.~\cite{papiro29},
\begin{eqnarray}
C_2(M_W) & = & -\frac{\lambda_{sb}^c}{\lambda_{sb}^t} \,, \nonumber \\
C_{7\gamma}(M_W) & = & -\frac{1}{2} D_0'(x_t) + \Delta C_{7\gamma}(M_W) \,,
\nonumber \\
C_{8G}(M_W) & = & -\frac{1}{2} E_0'(x_t) + \Delta C_{8G}(M_W) \,, \nonumber \\
C_3(M_W) & = & \Delta C_3 (M_W) \,, \nonumber \\[0.1cm]
C_7(M_W) & = & \Delta C_7 (M_W) \,, \nonumber \\[0.1cm]
C_9(M_W) & = & \Delta C_9 (M_W) \,.
\label{ec:48}
\end{eqnarray}
The extra terms in Model I are straightforward to include,
\begin{eqnarray}
\Delta C_{7\gamma}^\mathrm{I}(M_W) & = & -\frac{1}{2} 
\frac{\lambda_{sb}^T}{\lambda_{sb}^t} D_0'(x_T) \,, \nonumber \\
\Delta C_{8G}^\mathrm{I}(M_W) & = & -\frac{1}{2} 
\frac{\lambda_{sb}^T}{\lambda_{sb}^t} E_0'(x_T) \,, \nonumber
\\[0.1cm]
\Delta C_3^\mathrm{I} (M_W) & = & 0 \,, \nonumber \\[0.1cm]
\Delta C_7^\mathrm{I} (M_W) & = & 0 \,, \nonumber \\[0.1cm]
\Delta C_9^\mathrm{I} (M_W) & = & 0 \,.
\label{ec:49}
\end{eqnarray}
In Model $\modii$ there are contributions from $Z$ penguins with one or two FCN
couplings, plus $H$ penguins and other terms originated by the non-unitarity of
$V$. The expressions read \cite{papiro45}
\begin{eqnarray}
\Delta C_{7\gamma}^\modii (M_W) & = & \frac{X_{sb}}{\lambda_{sb}^t} \left(
\frac{23}{36} + \xi_s^Z +\xi_b^Z \right)
+ \frac{X_{sB} \, X_{Bb}}{\lambda_{sb}^t} [\xi_B^Z(y_B) + \xi_B^H(w_B)] \,,
\nonumber \\
\Delta C_{8G}^\modii (M_W) & = & \frac{X_{sb}}{\lambda_{sb}^t} \left(
\frac{1}{3} - 3 \xi_s^Z - 3 \xi_b^Z \right)
-3 \frac{X_{sB} \, X_{Bb}}{\lambda_{sb}^t} [\xi_B^Z(y_B) + \xi_B^H(w_B)] \,,
\nonumber \\
\Delta C_3^\modii (M_W) & = & -\frac{1}{6} \frac{X_{sb}}{\lambda_{sb}^t}
\,, \nonumber \\
\Delta C_7^\modii (M_W) & = & -\frac{2}{3} s_W^2 
\frac{X_{sb}}{\lambda_{sb}^t} \,, \nonumber \\
\Delta C_9^\modii (M_W) & = & \frac{2}{3} (1-s_W^2)
\frac{X_{sb}}{\lambda_{sb}^t} \,,
\label{ec:50}
\end{eqnarray}
with $y_i=(m_i/M_Z)^2$, $w_i=(m_i/M_H)^2$. The functions $\xi$ are given in
Appendix~\ref{sec:ap2}. We have made the approximations $y_s=0$, $y_b=0$,
$m_s/m_b=0$, and a very small term proportional to $X_{sd} X_{db}$ has been
omitted in $\Delta C_{7\gamma}^\modii$ and $\Delta C_{8G}^\modii$. Note also
that $C_2(M_W) = 1$ in the SM, but not necessarily in Models I and $\modii$.
The RG evolution to a scale $\mu=5$ GeV gives \cite{papiro45,papiro46b}
\begin{eqnarray}
C_{7\gamma}(\mu) & = & -0.158 \, C_2(M_W) + 0.695 \, C_{7\gamma}(M_W) 
+ 0.085 \, C_{8G}(M_W) \nonumber \\
& & + 0.143 \, C_3(M_W) + 0.101 \, C_7(M_W) -0.036 \, C_9(M_W) \,.
\label{ec:51}
\end{eqnarray}
From this coefficient we get $R_\gamma = 2.62 \times 10^{-3}$ in the SM. In
order to incorporate NLO corrections we normalise our LO calculation to the NLO
value \cite{papiro47} with an {\em ad hoc} factor
\footnote{We obtain the factor $K_\gamma$ comparing our LO and the NLO
calculation of $R_\gamma$ in Ref.~\cite{papiro47} with a common set of
parameters.}
$K_\gamma = 1.12$, and we keep the normalising factor for the calculation of
$R_\gamma$ in Models I and $\modii$. This is adequate provided the nonstandard
contributions are small. The systematic error of this approximation is
estimated to be smaller than
$\sim K_\gamma(K_\gamma-1)[R_\gamma - R_\gamma(\mathrm{SM})]$, and
vanishes if the new physics terms scale with the same factor $K_\gamma$. We have
found that in practice this error is of order $O(10^{-5})$, and in the worst
case $2 \times 10^{-4}$, smaller than the uncertainties present in the LO and
NLO calculations. With this procedure and using
$\mathrm{Br}(b \to c e \bar \nu) = 0.102$ we obtain
$\mathrm{Br}(b \to s \gamma) = 3.34 \times 10^{-4}$, in very good agreement
with the world average $(3.3 \pm 0.4) \times 10^{-4}$
\cite{papiro48,papiro48b,papiro48c}. We take as theoretical
uncertainty the one quoted in Ref.~\cite{papiro47}, $0.33 \times 10^{-4}$.

\subsubsection{$B \to X_s l^+ l^-$}
The analysis of the decay $B \to X_s l^+ l^-$ is very similar to the previous
one of $B \to X_s \gamma$. Again, the process can be approximated by
$b \to s l^+ l^-$ and the quantity theoretically obtained is the differential
ratio
\begin{equation}
R_{ll}(\hat s) \equiv \frac{1}{\Gamma (b \to c e \bar \nu)}
\frac{d}{d \hat s} \Gamma (b \to s l^+ l^-) \,,
\label{ec:52}
\end{equation}
with $\hat s = (p_{l^+}+p_{l^-})^2/m_b^2$ the normalised invariant mass of the
lepton pair. The calculation at LO \cite{papiro49} involves two more operators
at the scale $M_W$, $Q_{9V}$ and $Q_{10A}$. Defining for convenience
$\tilde C_{9V}$, $\tilde C_{10A}$ by
\begin{equation}
C_{9V} = \frac{\alpha}{2\pi} \, \tilde C_{9V} \,,~~
C_{10A} = \frac{\alpha}{2\pi} \, \tilde C_{10A} \,,
\label{ec:53}
\end{equation}
the latter are
\begin{eqnarray}
\tilde C_{9V}(M_W) & = & -\frac{4}{9} \frac{\lambda_{sb}^c}{\lambda_{sb}^t}+
\frac{Y_0(x_t)}{s_W^2}-4 Z_0(x_t) + \Delta \tilde C_{9V}(M_W) \,, \nonumber \\
\tilde C_{10A}(M_W) & = & - \frac{Y_0(x_t)}{s_W^2} + \Delta \tilde
C_{10A}(M_W) \,.
\label{ec:54}
\end{eqnarray}
The extra terms in Model I are analogous to the ones corresponding to the top,
\begin{eqnarray}
\Delta \tilde C_{9V}^\mathrm{I}(M_W) & = & \frac{\lambda_{sb}^T}{\lambda_{sb}^t}
\left[ \frac{Y_0(x_T)}{s_W^2}-4 Z_0(x_T) \right] \,, \nonumber \\
\Delta \tilde C_{10A}^\mathrm{I}(M_W) & = & -
 \frac{\lambda_{sb}^T}{\lambda_{sb}^t} \frac{Y_0(x_T)}{s_W^2} \,,
\label{ec:55}
\end{eqnarray}
and in Model $\modii$ we have \cite{papiro50}
\begin{eqnarray}
\Delta \tilde C_{9V}^\modii (M_W) & = & \left( \frac{1}{s_W^2}-4 \right)
C_{U2Z} \frac{X_{sb}}{\lambda_{sb}^t} \,, \nonumber \\
\Delta \tilde C_{10A}^\modii (M_W) & = & - \frac{C_{U2Z}}{s_W^2}
\frac{X_{sb}}{\lambda_{sb}^t} \,.
\label{ec:56}
\end{eqnarray}
The RG evolution to a scale $\mu = 5$ GeV gives the coefficients of the relevant
operators,
\begin{eqnarray}
C_1(\mu) & = & -0.221 \, C_2(M_W) \,, \nonumber \\
C_2(\mu) & = & 1.093 \, C_2(M_W) \,, \nonumber \\
\tilde C_{9V}(\mu) & = & \tilde C_{9V}(M_W) + 1.838 \, C_2(M_W) \,, \nonumber \\
\tilde C_{10A}(\mu) & = & \tilde C_{10A}(M_W) \,,
\label{ec:57}
\end{eqnarray}
and $C_{7\gamma}$ as in the process $b \to s \gamma$. We define for brevity in
the notation an ``effective'' $\tilde C_{9V}$,
\begin{equation}
\tilde C_{9V}^\mathrm{eff}(\mu) = \tilde C_{9V}(\mu) 
 + g(z,\hat s) \left[ 3\, C_1(\mu) + C_2(\mu) \right] \,,
\label{ec:58}
\end{equation}
where the $\hat s$-dependent function $g$ is \cite{papiro49}
\begin{eqnarray}
g(z,\hat s) & = & -\frac{8}{9} \log z + \frac{8}{27} + 
\frac{16}{9} \frac{z^2}{\hat s}
 \nonumber \\
& & - \frac{2}{9} \sqrt{\left| 1-\frac{4z^2}{\hat s} \right|}
 \left( 2 + \frac{4z^2}{\hat s} \right)  \times 
\left\{ \mbox{\begin{tabular}{lcc}
$2 \arctan \frac{1}{\sqrt{\frac{4z^2}{\hat s}-1}}$ & ~ & $\hat s < 4 z^2$ 
\\[0.2cm]
$\log \left| 
\frac{1+\sqrt{1-\frac{4z^2}{\hat s}}}{1-\sqrt{1-\frac{4z^2}{\hat s}}}
\right|+ \pi i$
& ~ & $\hat s > 4 z^2$
\end{tabular}} \right. \,.~~
\label{ec:59}
\end{eqnarray}
Then, the differential ratio $R_{ll}$ is written as
\begin{eqnarray}
R_{ll}(\hat s) & = & \frac{\alpha^2}{4 \pi^2 f(z)} 
\frac{|\lambda_{sb}^t|^2}{|V_{cb}|^2} (1-\hat s)^2 \left[ (1+2 \hat s)
(|\tilde C_{9V}^\mathrm{eff}|^2 + |\tilde C_{10A}|^2) \right. \nonumber \\
& & \left. + 4 (1+2/ \hat s) |\tilde C_{7\gamma}|^2
+ 12 \, \mathrm{Re} \, C_{7\gamma}^* \tilde C_{9V}^\mathrm{eff}
\right] \,.
\label{ec:60}
\end{eqnarray}
The partial width $\mathrm{Br}(b \to s l^+ l^-)$ is derived integrating $\hat s$
from $4 m_l^2/m_b^2$ to one and multiplying by the experimental value of
$\mathrm{Br}(b \to c e \bar \nu)$. Within the SM we obtain
$\mathrm{Br}(b \to s e^+ e^-) = 7.3 \times 10^{-6}$,
$\mathrm{Br}(b \to s \mu^+ \mu^-) = 5.0 \times 10^{-6}$. These values are a
little sensitive to the precise value of $s_W^2$ used. We use, as throughout
the paper, the $\overline \mathrm{MS}$ definition. Experimentally, 
$\mathrm{Br}(b \to s \mu^+ \mu^-) = (8.9 \pm 2.7) \times 10^{-6}$ but for
electrons only an upper bound exists,
$\mathrm{Br}(b \to s e^+ e^-) < 11.0 \times 10^{-6}$ with a 90\% CL
\cite{papiro51}. The NLO values \cite{papiro52} are 10\% larger but in view of
the experimental errors it is not necessary to incorporate NLO corrections to
set limits on new physics. The theoretical uncertainties, including the possible
modification of the $C_0$ functions, do not have much importance either (in
contrast with the decay $b \to s \gamma$) and we do not take them into account
in the statistical analysis.

Despite the worse experimental precision, in Model $\modii$
$b \to s l^+ l^-$ sets a stronger limit on the FCN coupling $X_{sb}$ than
$b \to s \gamma$. This is understood because in this model the decay
$b \to s l^+ l^-$ can be mediated by tree-level diagrams involving $X_{sb}$,
while in the process $b \to s \gamma$ this vertex appears only in extra penguin
diagrams and unitarity corrections, of the same size as the SM contributions.
Both $l=e$ and $l=\mu$ have to be considered, as the former sets the best upper
bound and the latter provides a lower bound. In Model I it also gives a more
restrictive constraint than $b \to s \gamma$, but we still include the latter in
the fit.

\subsection{The parameter $\varepsilon'$}
This parameter measures direct CP violation in the kaon system (its definition
can be found for instance in Ref.~\cite{libro}). For several years the
experimental measurements have been inconclusive, but now the determination has
settled, with a present accuracy of $\sim 10$ \%. On the contrary, the
theoretical prediction is subject to large uncertainties. Instead of calculating
$\varepsilon'$ directly, we calculate
$\varepsilon'/\varepsilon \simeq \mathrm{Re} \, \varepsilon'/\varepsilon$,
using the simplified expression \cite{papiro42}
\begin{equation}
\frac{\varepsilon'}{\varepsilon} = F_{\varepsilon'}(x_t) \; \mathrm{Im}\,
\lambda^t_{sd} + \Delta_{\varepsilon'} \,,
\label{ec:44}
\end{equation}
with
\begin{eqnarray}
F_{\varepsilon'}(x_t) & = & P_0 + P_X X_0(x_t) + P_Y Y_0(x_t) + P_Z Z_0(x_t)
+ P_E E_0(x_t)
\label{ec:45}
\end{eqnarray}
and $\Delta_{\varepsilon'}$ representing the new physics contribution. The
factors multiplying the Inami-Lim functions are
\begin{eqnarray}
P_0 & = & -3.167 + 12.409 \, B_6^{(1/2)} + 1.262 \, B_8^{(3/2)} \,, \nonumber \\
P_X & = & 0.540 + 0.023 \, B_6^{(1/2)} \,, \nonumber \\
P_Y & = & 0.387 + 0.088 \, B_6^{(1/2)} \,, \nonumber \\
P_Z & = & 0.474 - 0.017 \, B_6^{(1/2)} -10.186 \, B_8^{(3/2)} \,, \nonumber \\
P_E & = & 0.188 - 1.399 \, B_6^{(1/2)} + 0.459 \, B_8^{(3/2)} \,,
\end{eqnarray}
with $B_6^{(1/2)}$, $B_8^{(3/2)}$ non-perturbative parameters specified below.
The new contributions are
\begin{eqnarray}
\Delta^\mathrm{I}_{\varepsilon'} & = & F_{\varepsilon'}(x_T) \; \mathrm{Im}\,
\lambda^T_{sd} \nonumber \\
\Delta^\modii_{\varepsilon'} & = & C_{U2Z} (P_X+P_Y+P_Z) \, \mathrm{Im}\, X_{sd}
\label{ec:46}
\end{eqnarray}
In $\Delta^\mathrm{I}_{\varepsilon'}$ we approximate the $P$ coefficients in
$F_{\varepsilon'}(x_T)$ with the corresponding ones in $F_{\varepsilon'}(x_t)$.
From lattice or large $N_c$ calculations $B_6^{(1/2)} = 1.0 \pm 0.3$,
$B_8^{(3/2)} = 0.8 \pm 0.2$. Corrections accounting for final state
interactions \cite{papiro43,papiro43b} modify these figures to
$B_6^{(1/2)} = 1.55 \pm 0.5$, $B_8^{(3/2)} = 0.7 \pm 0.2$ yielding 
$\varepsilon'/\varepsilon = (1.64 \pm 0.70) \times 10^{-3}$ in good agreement
with the world average $(1.72 \pm 0.18) \times 10^{-3}$
\cite{papiro2a,papiro2b}. With large $N_c$ expansions at NLO very similar
results are obtained \cite{papiro43c}. Notice that the large theoretical error
in the $B$ parameters partially takes into account the different values from
different schemes. These uncertainties, together with cancellations among terms,
bring about a large uncertainty in the prediction. In spite of this fact
$\varepsilon'/\varepsilon$ is very useful to constrain the imaginary parts of
$\lambda^t_{sd}$, $\lambda^T_{sd}$ and $X_{sd}$ \cite{papiro44}.

\subsection{Summary}
The combined effect of the low energy constraints from $K$ and $B$ physics is to
disallow large cancellations and ``fine tuning'' of parameters to some extent.
As emphasised at the beginning of this Section, the various observables depend
on the CKM angles $V_{td}$, $V_{ts}$, $V_{tb}$ and the new physics parameters
in different functional forms. Therefore, if theoretical and experimental
precision were far better the parameter space would be constrained to a narrow
window around the SM values, and perhaps other possible regions allowed by
cancellations. However, as can be seen in Table~\ref{tab:5}, present theoretical
and experimental precision allow for relatively large contributions of the new
physics in Models I and $\modii$, with large deviations in some observables.

\begin{table}[htb]
\begin{center}
\begin{tabular}{ccc}
\hline
\hline
 & SM prediction & Exp. value \\
\hline
$\varepsilon$ & $2.18 \times 10^{-3}$ &
  $(2.282 \pm 0.017) \times 10^{-3}$ \\
$|\delta m_B|$ & $0.49$ &
  $0.489 \pm 0.008$ \\
$a_{\psi K_S}$ & $0.71$ & $0.734 \pm 0.054$ \\
$|\delta m_{B_s}|$ & $17.6$ &
  $> 13.1$ (95\%) \\
$|\delta m_D|$ & $\sim 10^{-5}$ & $< 0.07$ (95\%) \\
$\mathrm{Br}(K^+ \to \pi^+ \nu \bar \nu)$ & $6.4 \times 10^{-11}$ &
  $[3.2-48] \times 10^{-11}$ (90\%) \\
$\mathrm{Br}(K_L \to \mu^+ \mu^-)_\mathrm{SD}$ & $6.6 \times 10^{-10}$ &
  $< 1.9 \times 10^{-9}$ (90\%) \\
$\mathrm{Br}(b \to s \gamma)$ & $3.34 \times 10^{-4}$ &
  $(3.3 \pm 0.4) \times 10^{-4}$ \\
$\mathrm{Br}(b \to s e^+ e^-)$ & $7.3 \times 10^{-6}$ &
  $< 11.0 \times 10^{-6}$ (90\%) \\
$\mathrm{Br}(b \to s \mu^+ \mu^-)$ & $5.0 \times 10^{-6}$ &
  $(8.9 \pm 2.7) \times 10^{-6}$ \\
$\varepsilon'/\varepsilon$ & $1.64 \times 10^{-3}$ &
  $(1.72 \pm 0.18) \times 10^{-3}$ \\
\hline
\hline
\end{tabular}
\caption{Experimental values of the low energy observables used in
the fits, together with the SM calculations with the parameters in
Appendix~\ref{sec:ap1}. The theoretical errors can be found in the text.
The mass differences are in ps$^{-1}$.
\label{tab:5}}
\end{center}
\end{table}

Other potential restrictions on these models have been explored: the rare
decays $B \to l^+ l^-$, $B_s \to l^+ l^-$ and $B \to X_s \nu \bar \nu$. With
present experimental precision they do not provide additional constraints, nor
the predictions for their rates differ substantially from SM expectations. An
important exception is $\mathrm{Br}(K_L \to \pi^0 \nu \bar \nu)$. This decay
mode does not provide a constraint yet, but in Models I and $\modii$ it can have
a branching ratio much larger than in the SM. Its analysis is postponed to
Section~\ref{sec:7}.

\section{Other constraints}
\label{sec:6}
The diagonal couplings of the $u,d$ quarks to the $Z$ boson are extracted from
neutrino-nucleon scattering processes, atomic parity violation and the SLAC
polarised electron experiment (see Ref.~\cite{papiro3} and references therein
for a more extensive discussion). The values of $c_{L,R}^u$, $c_{L,R}^d$ derived
from $\nu N$ neutral processes have a large non-Gaussian correlation, and for
the fit it is convenient to use instead
\begin{eqnarray}
g_L^2 & = & \frac{1}{4} \left[ (c_L^u)^2 + (c_L^d)^2 \right] \,, \nonumber \\
g_R^2 & = & \frac{1}{4} \left[ (c_R^u)^2 + (c_R^d)^2 \right] \,, \nonumber \\
\theta_L & = & \arctan \frac{c_L^u}{c_L^d} \,, \nonumber \\
\theta_R & = & \arctan \frac{c_R^u}{c_R^d} \,.
\label{ec:65}
\end{eqnarray}
The SM predictions for these parameters, including radiative corrections
\cite{papiro3} and using the $\overline \mathrm{MS}$ definition of $s_W^2$,
are collected in Table~\ref{tab:6}, together with their experimental values. The
correlation matrix is in Table~\ref{tab:7}.

\begin{table}[htb]
\begin{center}
\begin{tabular}{lccc}
\hline
\hline
& SM & Experimental \\[-0.1cm]
& prediction & measurement & Error \\
\hline
$g_L^2$   & $0.3038$ & $0.3020$ & $0.0019$ \\
$g_R^2$   & $0.0300$ & $0.0315$ & $0.0016$ \\
$\theta_L$ & $2.4630$ & $2.50$ & $0.034$ \\
$\theta_R$ & $5.1765$ & $4.58$ & $^{+0.40}_{-0.27}$ \\[0.1cm]
\hline
\hline
\end{tabular}
\end{center}
\caption{SM calculation of $g_L^2$, $g_R^2$, $\theta_L$, $\theta_R$
and experimental values.
\label{tab:6}}
\end{table}

\begin{table}[htb]
\begin{center}
\begin{tabular}{lcccc}
\hline
\hline
 & $g_L^2$ & $g_R^2$ & $\theta_L$ & $\theta_R$ \\[0.1cm]
\hline
$g_L^2$ & $1$ & $0.32$ & $-0.39$ & $\sim 0$  \\
$g_R^2$ & $0.32$ & $1$ & $-0.10$ & $\sim 0$  \\
$\theta_L$ & $-0.39$ & $-0.10$ & $1$ & $0.27$  \\
$\theta_R$ & $\sim 0$ & $\sim 0$ & $0.27$ & $1$  \\
\hline
\hline
\end{tabular}
\end{center}
\caption{Correlation matrix for the parameters
$g_L^2$, $g_R^2$, $\theta_L$, $\theta_R$.
\label{tab:7}}
\end{table}

The interactions involved in atomic parity violation and the SLAC polarised
electron experiment can be parameterised with the effective Lagrangian
\begin{equation}
\mathcal{L} = -\frac{G_F}{\sqrt 2} \sum_{i=u,d} \left[
C_{1i} \bar e \gamma_\mu \gamma^5 e \, \bar q_i \gamma^\mu q_i
+ C_{2i} \bar e \gamma_\mu e \, \bar q_i \gamma^\mu \gamma^5 q_i \right]
\label{ec:66}
\end{equation}
plus a QED contribution. We are not considering mixing of the leptons, and hence
the coefficients $C_{1i}$, $C_{2i}$ are at tree level \cite{papiro11c}
\begin{eqnarray}
C_{1u} & = & - \left( \frac{X_{uu}}{2} - \frac{4}{3} s_W^2 \right) \,,
\nonumber \\
C_{1d} & = & - \left(- \frac{X_{dd}}{2} + \frac{2}{3} s_W^2 \right) \,,
\nonumber \\
C_{2u} & = & X_{uu} \left( -\frac{1}{2}+2 s_W^2 \right) \,,
\nonumber \\
C_{2d} & = & -X_{dd} \left( -\frac{1}{2}+2 s_W^2 \right) \,.
\label{ec:67}
\end{eqnarray}
The parameters $C_{1u}$ and $C_{1d}$ can be extracted from atomic parity
violation measurements. The combination
$\tilde C_2 \equiv C_{2u}-C_{2d}/2$ is obtained in the polarised electron
experiment. In Table~\ref{tab:8} we quote the SM predictions of $C_{1u}$,
$C_{1d}$, $\tilde C_2$, including the radiative corrections to
Eqs.~(\ref{ec:67}), and the experimental values. The correlation matrix is in
Table~\ref{tab:9}.

\begin{table}[htb]
\begin{center}
\begin{tabular}{lccc}
\hline
\hline
& SM & Experimental \\[-0.1cm]
& prediction & measurement & Error \\
\hline
$C_{1u}$ & $-0.1886$ & $-0.209$ & $0.041$ \\
$C_{1d}$ & $0.3413$ & $0.358$ & $0.037$ \\
$\tilde C_2$ & $-0.0492$ & $-0.04$ & $0.12$ \\
\hline
\hline
\end{tabular}
\end{center}
\caption{Experimental values and SM calculations of the parameters
$C_{1u}$, $C_{1d}$ in the Lagrangian in Eq.~(\ref{ec:66}) and the combination
$\tilde C_2 = C_{2u}-C_{2d}/2$.
\label{tab:8}}
\end{table}

\begin{table}[htb]
\begin{center}
\begin{tabular}{lccc}
\hline
\hline
 & $C_{1u}$ & $C_{1d}$ & $\tilde C_2$  \\[0.1cm]
\hline
$C_{1u}$ & $1$ & $-0.9996$ & $-0.78$  \\
$C_{1d}$ & $-0.9996$ & $1$ & $0.78$   \\
$\tilde C_2$ & $-0.78$ & $0.78$ & $1$  \\
\hline
\hline
\end{tabular}
\end{center}
\caption{Correlation matrix for the parameters
$C_{1u}$, $C_{1d}$ and $\tilde C_2 = C_{2u}-C_{2d}/2$.
\label{tab:9}}
\end{table}

The leading effect of mixing with singlets is a decrease of $X_{uu}$ (in Model
I) and $X_{dd}$ (in Model $\modii$), which reduces $g_L^2$ and the modulus of
$\tilde C_2$ in both cases, and also the modulus of $C_{1u}$ or $C_{1d}$,
depending on the model considered. The angle $\theta_L$ grows when the up quark
mixes with a singlet and decreases when the mixing corresponds to the down
quark. The right-handed couplings are not affected by mixing with singlets, and
thus $g_R^2$ and $\theta_R$ remain equal to their SM values. However, they have
to be included in the fit because of the experimental correlation.  Another
consequence of the mixing may possibly be the modification of the small
radiative corrections to Eqs.~(\ref{ec:65},\ref{ec:67}). We neglect this
subleading effect and assume that the corrections remain with their SM values.

\section{Some observables from $K$ and $B$ physics with large new effects}
\label{sec:7}
There is a large number of observables of interest that will be tested in
experiments under way and for which our models lead to departures from the SM.
Necessarily, our study is not complete and we pass over many relevant processes
that deserve further attention. We discuss for illustration the CP-violating
decay $K^0 \to \pi^0 \nu \bar \nu$ and the time-dependent CP asymmetry in
$B_s^0 \to D_s^+ D_s^-$.

\subsection{The decay $K^0 \to \pi^0 \nu \bar \nu$}
This decay is closely related to $K^+ \to \pi^+ \nu \bar \nu$, but its
detection is much more difficult. At present it is still unobserved, and the
90\% CL limit on this decay mode is $\mathrm{Br}(K_L \to \pi^0 \nu \bar \nu)
\leq 5.9 \times 10^{-7}$ \cite{papiro61}. It is calculated as
\begin{eqnarray}
\mathrm{Br}(K_L \to \pi^0 \nu \bar \nu) & = & 
r_{K_L} \frac{\tau_{K_L}}{\tau_{K^+}}
\mathrm{Br}(K^+ \to \pi^0 e^+ \bar \nu)
\;\frac{3 \alpha^2}{2 \pi^2 s_W^4 |V_{us}|^2} 
\nonumber \\
& & \times \left[ \eta_t^X X_0(x_t) \, \mathrm{Im} \,
\lambda_{sd}^t + \mathrm{Im} \, \Delta_{K^+} \right]^2 \,,
\end{eqnarray}
with the corresponding isospin breaking correction $r_{K_L} = 0.944$.
The SM prediction for this partial width is $(2.3 \pm 0.17) \times 10^{-11}$,
one third of the value for $\mathrm{Br}(K^+ \to \pi^+ \nu \bar \nu)$.
As only the imaginary parts of the CKM products enter in the expression for
$\mathrm{Br}(K_L \to \pi^0 \nu \bar \nu)$, it is possible to have this decay
rate much larger while keeping $\mathrm{Br}(K^+ \to \pi^+ \nu \bar \nu)$
in agreement with experiment. However, there is a model-independent limit
\begin{equation}
\frac{\mathrm{Br}(K_L \to \pi^0 \nu \bar \nu)}{\mathrm{Br}(K^+ \to \pi^+ \nu
  \bar \nu)} \leq 4.376 \,
\label{ec:n1}
\end{equation}
that holds provided lepton flavour is conserved \cite{papiro62}. We will find
later that in our models this bound can be saturated, leading to a increase in
$\mathrm{Br}(K_L \to \pi^0 \nu \bar \nu)$ of an order of magnitude, even keeping
$\mathrm{Br}(K^+ \to \pi^+ \nu \bar \nu)$ at its SM value.

\subsection{The CP asymmetry in $B_s^0 \to D_s^+ D_s^-$.}
The ``gold plated'' decay $B_s^0 \to D_s^+ D_s^-$ is mediated by the
quark-level transition $\bar b \to \bar c c \bar s$. This transition is
dominated by a single tree-level amplitude $A \propto V_{cb}^* V_{cs}$ and it is
then free of hadronic uncertainties \cite{papiro64}. The time-dependent CP
asymmetry is written as  $a_{D_s^+ D_s^-} = \mathrm{Im}\,\lambda_{D_s^+ D_s^-}$,
with \cite{papiro63}
\begin{equation}
\lambda_{D_s^+ D_s^-} = \frac{(M_{12}^{B_s})^*}{|M_{12}^{B_s}|} \,
\frac{V_{cb} V_{cs}^*}{V_{cb}^* V_{cs}}
\end{equation}
and also
\begin{equation}
a_{D_s^+ D_s^-} = \sin (2\zeta-2\theta_{B_s}) \,,
\end{equation}
with
\begin{equation}
\zeta = \arg \left[ -\frac{V_{cb} V_{cs}^*}{V_{tb} V_{ts}^*} \right] \,.
\end{equation}
and $\theta_{B_s}$ parameterising the deviation of the phase of $M_{12}^{B_s}$
with respect to its SM value,
\begin{equation}
2\,\theta_{B_s} = \arg \frac{M_{12}^{B_s}}{(M_{12}^{B_s})_\mathrm{SM}} \,.
\end{equation}
In the SM $\zeta \simeq 0$, so the asymmetry $a_{D_s^+ D_s^-}$ is predicted to
be very
small ($a_{D_s^+ D_s^-} \simeq 0.03$ with the parameters for the CKM matrix in
Appendix~\ref{sec:ap1}). Therefore, its measurement offers a good opportunity to
probe new physics, which may manifest itself if a nonzero value is observed
\cite{papiro63}.

Another possible final state given by the same quark-level transition
is $\psi \phi$. This state has a clean experimental
signature at hadron colliders, $\psi \to l^+ l^-$ and $\phi \to K^+ K^-$,
providing better chances to measure $\sin (2\zeta-2\theta_{B_s})$ at
Tevatron \cite{referee1}. However, in this case both particles $\psi$
and $\phi$ have spin 1, and then the orbital angular momentum is not fixed.
(In $B_s^0 \to D_s^+ D_s^-$ the
two $D$ mesons have spin 0, and therefore they are produced in a $l=0$ CP-even
state.) The $\psi \phi$ are produced in an admixture of CP-even and CP-odd
states, which can be disentangled with an analysis of the angular distribution
of their decay products $l^+ l^-$, $K^+ K^-$ \cite{referee2}. We will loosely
refer to the CP asymmetries containing the phase $(2\zeta-2\theta_{B_s})$ as
$a_{D_s^+ D_s^-}$, understanding that this includes the asymmetries
corresponding to final states
$D_s^+ D_s^-$, $\psi \phi$, $D_s^{*+} D_s^{*-}$, etc.
Finally, it is worthwhile mentioning that $\cos (2\zeta-2\theta_{B_s})$ can
also be
measured, on condition that the width difference between the two mass
eigenstates is sizeable \cite{referee3}. This can be done without the need of
tagging the initial state ($B_s^0$ or $\bar B_s^0$) and provides an independent
measurement of this important phase.

\section{Results}
\label{sec:8}
We are more interested in the departures from SM predictions originated
by mixing with exotic singlets than in finding the best fit to all experimental
data. Therefore we must specify the criteria for what we will consider as
agreement of these models with experiment, and of course our predictions depend
on this choice. We require: ({\em i\/}) individual agreement of observables with
data, and ({\em ii\/}) that the joint $\chi^2$ of the observables, divided in
subsets, is not much worse than the $\chi^2$ of these subsets in the SM. For the
first condition, the number of standard deviations allowed in a single
observable is similar to the departure already present in the SM. The second
condition consists in requiring that the $\chi^2$ of a group of variables is
smaller or equal to the SM $\chi^2$ value increased by a quantity numerically
equal to the number of variables in the group. In average, this condition means
admitting a $1\sigma$ deviation for a variable which in the SM coincides with
the experimental measurement, an extra departure of $0.41\sigma$ for a variable
which is at $1\sigma$ within the SM, or $0.24\sigma$ for a variable which is
already at $2\sigma$ in the SM. This second condition is in practice much
stronger than the first one. These criteria are best explained by enumerating
them:
\begin{itemize}
\item The moduli of the CKM angles $V_{ud}$, $V_{us}$, $V_{ub}$, $V_{cd}$,
$V_{cs}$, $V_{cb}$ can be at most $2\sigma$ away from the figures in
Table~\ref{tab:0}. The sum of the $\chi^2$ must be smaller than the SM result
plus six.
\item The predictions for $R_b$, $R_c$, $A_\mathrm{FB}^{0,b}$,
$A_\mathrm{FB}^{0,c}$, $\mathcal{A}_b$, $\mathcal{A}_c$ may be up to $3\sigma$
away from the central values in Table~\ref{tab:1}. We allow larger departures
than in the previous case because the SM prediction of $A_\mathrm{FB}^{0,b}$
is almost $3\sigma$ from the experimental measurement. We also require that the
$\chi^2$ (calculated with the correlation matrix in Table~\ref{tab:2}) is
smaller than the SM result plus four (in Model I) or plus three (in Model
$\modii$). The number of variables in this subset that effectively change with
the mixing are four and three in Models I and $\modii$, respectively.
\item The contributions to the oblique parameters $S$, $T$, $U$ from new physics
have to be within $2\sigma$ of the values in Table~\ref{tab:3}. The sum of the
$\chi^2$ of the three variables must be smaller than the SM $\chi^2$ value plus
three.
\item The observables $\varepsilon$, $|\delta m_B|$ and $a_{\psi K_S}$ are
allowed to move within $2.5\sigma$ of the numbers in Table~\ref{tab:5}. The
total $\chi^2$ has to be smaller than the SM value plus three.
\item The branching fractions for $b \to s \gamma$ and $b \to s \mu^+ \mu^-$ are
required to agree with the experimental figures in Table~\ref{tab:5} within
$2\sigma$, and their $\chi^2$ has to be smaller than the SM result plus two.
\item The departure of $\varepsilon'/\varepsilon$ from the experimental
measurement in Table~\ref{tab:5} can be at most $1\sigma$ larger than the
departure within the SM.
\item The parameters $g_L^2$, $g_R^2$, $\theta_L$, $\theta_R$, $C_{1u}$,
$C_{1d}$ and $\tilde C_2$ have to be within $2\sigma$ of the central values in
Tables~\ref{tab:6} and \ref{tab:8}. Their $\chi^2$ computed with the correlation
matrices in Tables~\ref{tab:7} and \ref{tab:9} is required to be smaller than
the SM value plus five ($g_R$ and $\theta_R$ are not affected by the mixing in
these models).
\end{itemize}
The observables $\varepsilon$, $|\delta m_B|$, $\mathrm{Br}(b \to s \gamma)$
and $\varepsilon'/\varepsilon$ have large theoretical errors that are of similar
magnitude as the experimental ones. In the comparison of these observables with
experiment we use the prescription explained in Appendix~\ref{sec:ap3},
assuming that the theoretical errors are Gaussian. For $|\delta m_D|$, 
$\mathrm{Br}(K^+ \to \pi^+ \nu \bar \nu)$,
$\mathrm{Br}(K_L \to \mu^+ \mu^-)_\mathrm{SD}$ and
$\mathrm{Br}(b \to s e^+ e^-)$ we require that the predictions are in the
experimental intervals quoted in Table~\ref{tab:5} (the upper limit of
$|\delta m_D|$ in the literature has a 95\% CL, instead of the more common one
of 90\%). We also set the condition $|\delta m_{B_s}/\delta m_B| \geq 26.7$,
rather than $|\delta m_{B_s}| > 13.1$ ps$^{-1}$, to avoid theoretical
uncertainties. With all these restrictions we explore the parameter space to
find the interval of variation of charged current top couplings,
flavour-diagonal and flavour-changing $Z$ couplings and the observables
introduced in last Section. We discuss the results for Models I and $\modii$
separately.

\subsection{Mixing with an up singlet}
One fundamental parameter in Model I is the mass of the new quark $m_T$. We find
that for low $m_T$ the effects of mixing can be huge, with $V_{td}$, $V_{ts}$
and $V_{tb}$ very different from the SM predictions. In this scenario the
effects of mixing on $R_b$ and oblique corrections almost cancel, while the new
quark can virtually take the place of the top in reproducing the experimental
values of the meson observables analysed. This is viable because for small $m_T$
the Inami-Lim functions for the top and the new quark are alike. As $m_T$ grows
this possibility disappears and the dominant contributions come from the top,
but still significant departures from the SM can be found. In our analysis we
have checked that for the values and plots shown the Yukawa couplings remain
perturbative. The decoupling limit is not reached in any of the cases
considered.

\begin{figure}[htb]
\begin{center}
\mbox{\epsfig{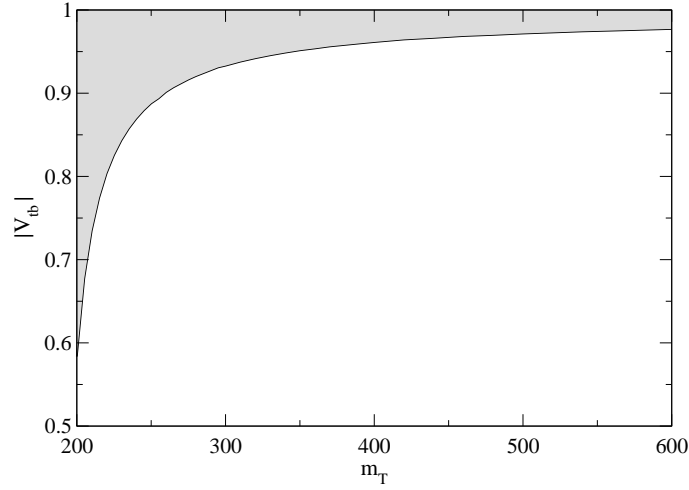}}
\end{center}
\caption{Allowed values of $|V_{tb}|$ (shaded area) in Model I, as a function of
the mass of the new quark.
\label{fig:vtb} }
\end{figure}

\begin{figure}[htb]
\begin{center}
\mbox{\epsfig{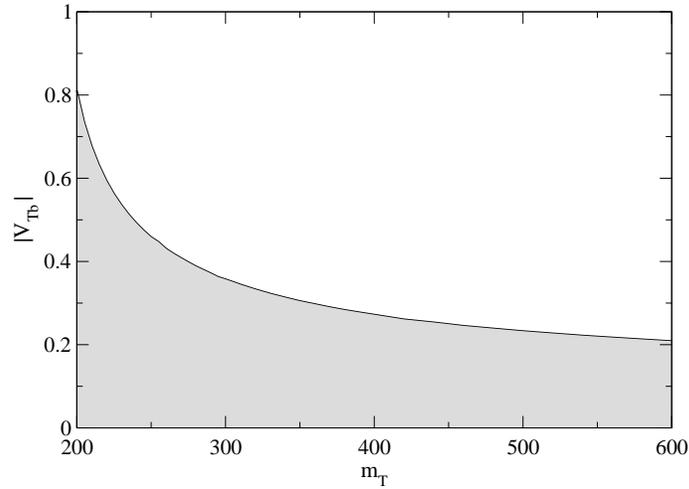}}
\end{center}
\caption{Allowed values of the coupling $|V_{Tb}|$ of the new quark
(shaded area) in Model I, as a function of its mass.
\label{fig:vtb2} }
\end{figure}

One of the most striking results in Model I is the deviation of $|V_{tb}|$ from
unity (see Fig.~\ref{fig:vtb}). The modulus of $V_{tb}$ is determined by the
coupling $V_{Tb}$ in Fig.~\ref{fig:vtb2}, and the latter is bounded by the $T$
parameter, as it was seen in Section~\ref{sec:4}. (The dependence on only one
observable leads to the very simple behaviour of the curves in
Figs.~\ref{fig:vtb} and\ref{fig:vtb2}.) For $m_T = 200$ GeV, $|V_{tb}|$ can be
as small as $0.58$. The lower limit on $|V_{tb}|$ grows with $m_T$, but even for
$m_T = 600$ GeV it is $|V_{tb}| \geq 0.977$, substantially different from the SM
prediction $|V_{tb}| = 0.999$. Although sizeable and theoretically very
important, this 2\% difference is difficult to detect experimentally at LHC,
which is expected to measure the size of $V_{tb}$ with a precision of $\pm 0.07$
\cite{papiro9a}.

\begin{figure}[htb]
\begin{center}
\mbox{\epsfig{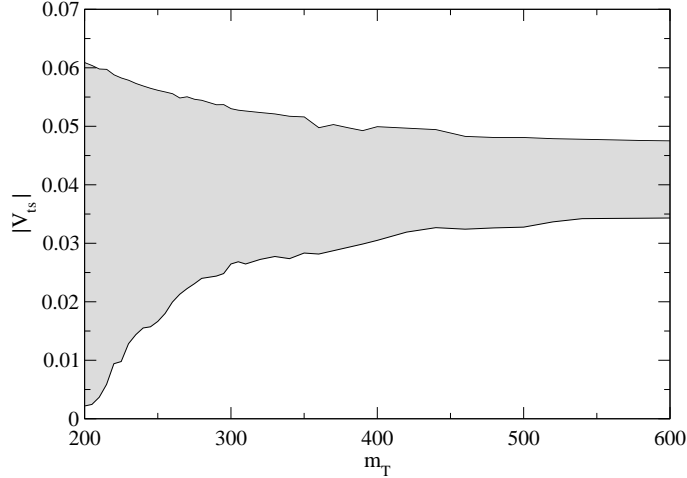}}
\end{center}
\caption{Allowed values of $|V_{ts}|$ (shaded area) in Model I, as a function of
the mass of the new quark.
\label{fig:vts} }
\end{figure}

\begin{figure}[htb]
\begin{center}
\mbox{\epsfig{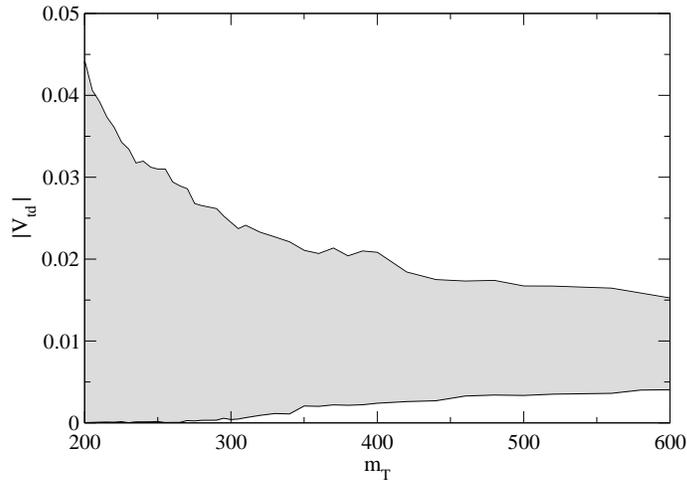}}
\end{center}
\caption{Allowed values of $|V_{td}|$ (shaded area) in Model I, as a function of
the mass of the new quark.
\label{fig:vtd} }
\end{figure}

\begin{figure}[htb]
\begin{center}
\mbox{\epsfig{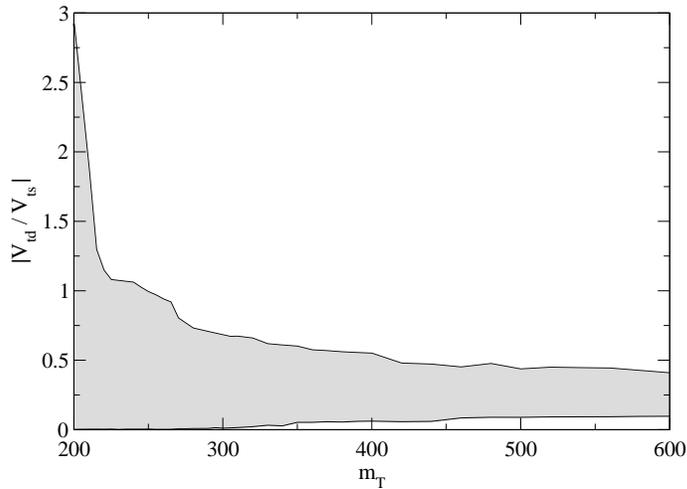}}
\end{center}
\caption{Allowed values of the ratio $|V_{td}/V_{ts}|$ (shaded area) in Model
I, as a function of the mass of the new quark.
\label{fig:vr} }
\end{figure}

The top charged-current couplings $V_{ts}$ and $V_{td}$ can be very different
from SM expectations as well. In the SM $3 \times 3$ CKM unitarity fixes
$|V_{ts}| \simeq |V_{cb}|$. In Model I $|V_{ts}|$ can be between $0.002$ and
$0.061$ for $m_T = 200$ GeV (see Fig.~\ref{fig:vts}). The allowed interval
narrows as $m_T$ increases, and for $m_T = 600$ GeV the interval is essentially
the same as in the SM. The range of variation of $V_{td}$ is also considerably
greater than in the SM (see Fig.~\ref{fig:vtd}). For $m_T \leq 300$ GeV $V_{td}$
can be almost zero (and in this case the $T$ quark would account for the
measured values of $K$ and $B$ observables), or even larger than $V_{ts}$, as
can be seen in Fig.~\ref{fig:vr}. Again, for heavier $T$ the permitted interval
decreases and for $m_T = 600$ GeV it is practically the same interval as in the
SM. We remark that the curves in Figs.~\ref{fig:vts}-\ref{fig:vr} giving the
upper and lower bounds arise from the various restrictions discussed in
Sections~\ref{sec:3}-\ref{sec:6}, especially those regarding meson observables,
thus their complicated behaviour should not be surprising. We do not claim that
the blank regions in these three figures are excluded. The quoted allowed limits
might be wider if some delicate cancellation not found in the numerical analysis
allows a small region in parameter space with $V_{td}$, $V_{ts}$ or their ratio
outside the shaded areas. 

In contrast with the former, the intervals for CKM mixing angles $V_{Td}$,
$V_{Ts}$ do not show a pronounced decrease with $m_T$. $V_{Td}$ can be in the
interval $0 \leq |V_{Td}| \leq 0.05$ for the $m_T$ values studied, and the
maximum size of $|V_{Ts}|$ decreases from $0.06$ for $m_T = 200$ to $0.05$ for
$m_T = 600$ GeV.

\begin{figure}[htb]
\begin{center}
\mbox{\epsfig{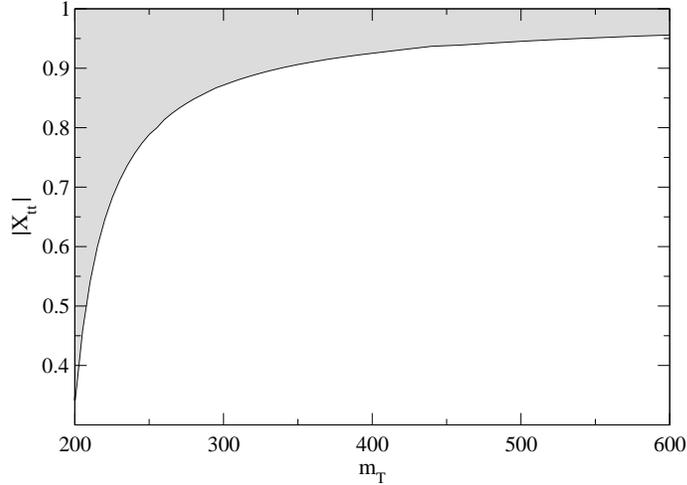}}
\end{center}
\caption{Allowed values of $|X_{tt}|$ (shaded area) in Model I, as a function of
the mass of the new quark.
\label{fig:xtt} }
\end{figure}

The counterpart of the departure from the SM prediction $|V_{tb}| = 0.999$ is
the decrease of the $Ztt$ coupling. Within the SM, the isospin-related term
$X_{tt}$ equals one by the GIM mechanism, while in Model I the GIM breaking
originated by mixing with a singlet reduces its magnitude. The modulus of
$X_{tt}$, as well as $V_{tb}$, is determined by the parameter $V_{Tb}$ and hence
its possible size is dictated only by the $T$ parameter. The interval allowed
for $X_{tt}$ is plotted in Fig.~\ref{fig:xtt}, where we observe that for
$m_T = 200$ GeV it reaches down to $X_{tt}=0.34$. The lower limit of the
interval grows with $m_T$ and is approximately $X_{tt}=0.96$ for $m_T=600$ GeV.
The $Ztt$ coupling will be precisely measured in $t \bar t$ production at TESLA.
With a CM energy of 500 GeV and an integrated luminosity of 300 fb$^{-1}$, 34800
top pairs are expected to be collected at the detector in the semileptonic
channel $l\nu jjjj$, with $l$ an electron or a muon. The estimated precision in
the determination of $X_{tt}$ with this channel alone is of $0.02$. Then, even
with $m_T = 600$ GeV a $2\sigma$ effect could be visible.

\begin{figure}[htb]
\begin{center}
\mbox{\epsfig{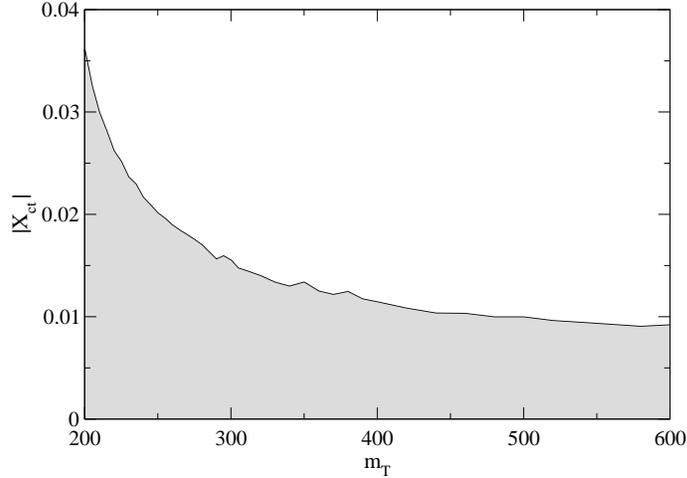}}
\end{center}
\caption{Allowed values of the FCN coupling $|X_{ct}|$ (shaded area)
in Model I, as a function of the mass of the new quark.
\label{fig:xct} }
\end{figure}

FCN couplings are perhaps the most conspicuous manifestation of mixing with
quark singlets, and offer another excellent place to search for new physics.In
the SM they vanish at tree-level by the GIM mechanism, and the effective
vertices generated at one loop are very small as a consequence of the GIM
suppression \cite{papiro65}. This results in a negligible branching ratio
$\mathrm{Br}(t \to Zc) \sim 10^{-14}$ within the SM. In Model I the FCN coupling
$X_{ct}$ can be sizeable \cite{papiro16}, leading to top decays $t \to Zc$
\cite{papiro66}, $Zt$ production at LHC \cite{papiro67,papiro68} and single top
production at linear colliders \cite{papiro69,papiro70,papiro71,papiro72}. For
$m_T \sim m_t$ the new contributions to meson observables involving $T$ diagrams
are small, and this FCN coupling can be relatively large, $|X_{ct}| = 0.036$
(see Fig.~\ref{fig:xct})
\footnote{The reduction with respect to the number quoted in
Ref.~\cite{papiro16} is mainly due to the improved limit on $|\delta m_D|$.}.
A coupling of this size yields a branching ratio
$\mathrm{Br}(t \to Zc) = 6.0 \times 10^{-4}$ (nine orders of magnitude above the
SM prediction) that would be seen at LHC with $18\sigma$ statistical
significance in top decays and $4.6\sigma$ in $Zt$ production (with an
integrated luminosity of 100 fb$^{-1}$), and at TESLA with $8.2\sigma$
significance in single top production (with 300 fb$^{-1}$). For larger $m_T$,
the contributions of the $T$ quark to meson observables (in particular to
$K^+ \to \pi^+ \nu \bar \nu$ and the short-distance part of
$K_L \to \mu^+ \mu^-$) decrease monotonically the upper limit on $|X_{ct}|$,
with some very small local ``enhancements'' that can be observed in
Fig.~\ref{fig:xct}. For $T$ very heavy there is still the possibility of
$|X_{ct}| = 0.009$, giving  $\mathrm{Br}(t \to Zc) = 3.8 \times 10^{-5}$, which
would have a $1.2\sigma$ significance in top decay processes at LHC. 

In Model I the $X_{ut}$ coupling can have the same size as $X_{ct}$. This
contrasts with other SM extensions (for instance, SUSY or two Higgs doublet
models) where observable FCN $tc$ vertices can be generated but $tu$ vertices
are suppressed. The observability of a $Ztu$ FCN coupling is the same, and even
better in the case of $Zt$ production processes at LHC. The coupling $X_{tT}$
between the top and the new mass eigenstate (which is a function of the
charged-current coupling $V_{Tb}$) can reach the maximum value permitted by the
model, $X_{tT} = 0.5$ for $m_T \leq 210$ GeV, descending slowly to a maximum of
$X_{tT} = 0.2$ when $m_T = 600$ GeV.

\begin{figure}[htb]
\begin{center}
\mbox{\epsfig{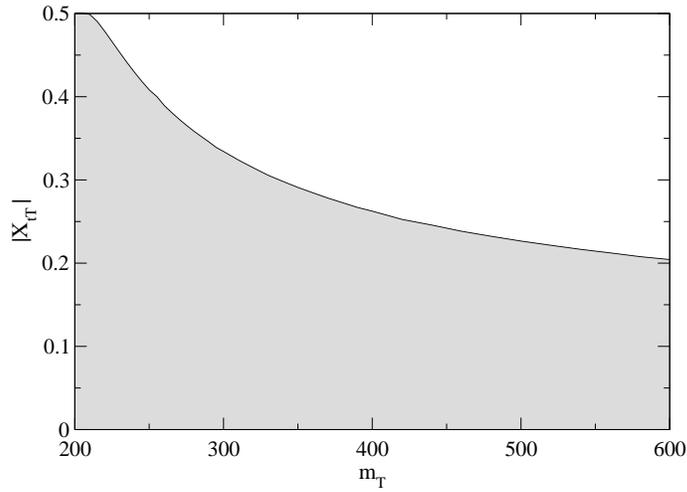}}
\end{center}
\caption{Allowed values of the coupling $|X_{tT}|$ of the new quark
(shaded area) in Model I, as a function of its mass.
\label{fig:xtt2} }
\end{figure}

\begin{figure}[htb]
\begin{center}
\mbox{\epsfig{file=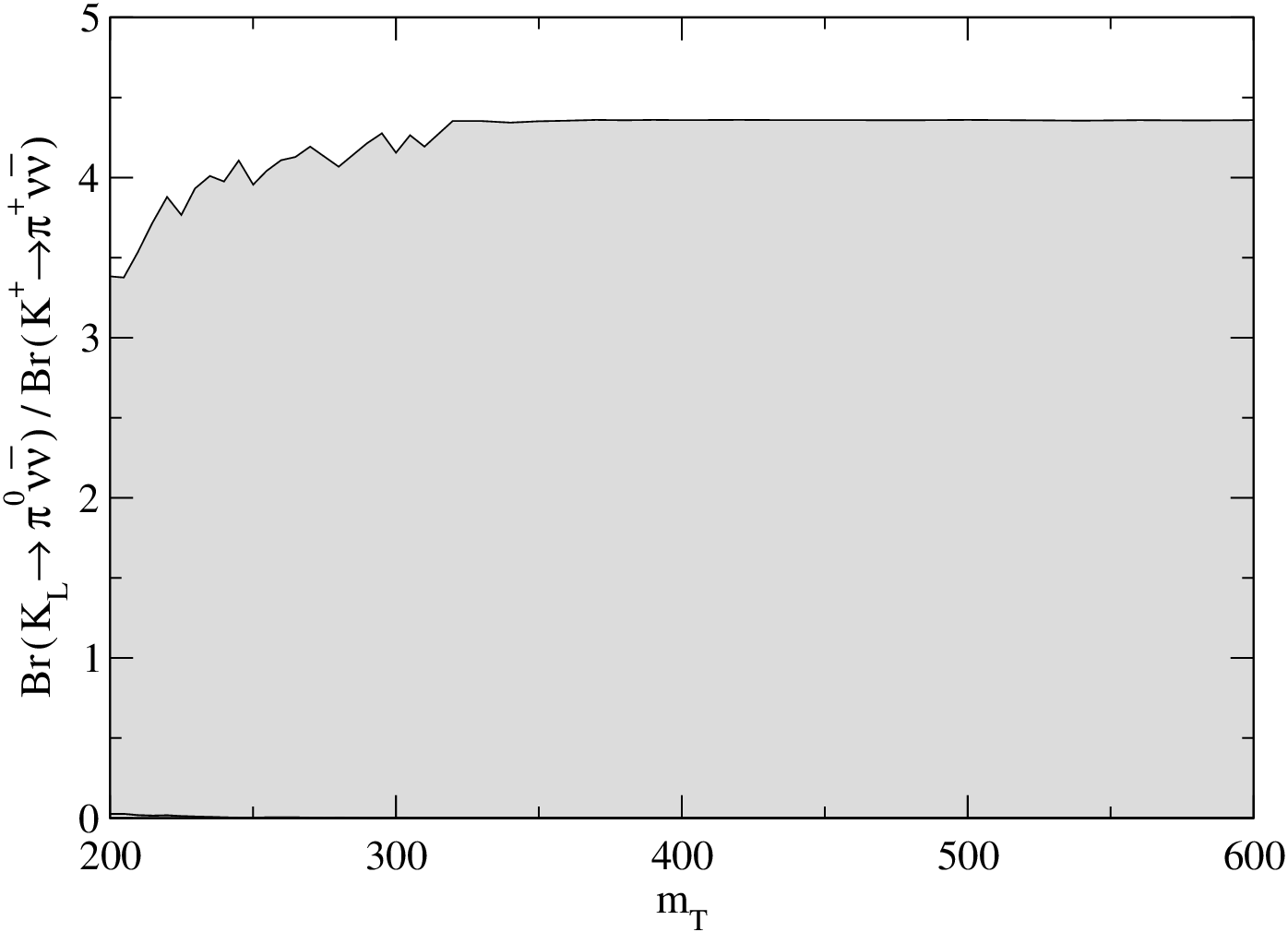,width=9cm,clip=}}
\end{center}
\caption{Range of variation of $\mathrm{Br}(K_L \to \pi^0 \nu \bar \nu)/
\mathrm{Br}(K^+ \to \pi^+ \nu \bar \nu)$ in Model I (shaded area).
\label{fig:brk} }
\end{figure}

The mixing with a new singlet may also give new effects in low energy
observables. The branching ratio of $K_L \to \pi^0 \nu \bar \nu$ can reach
$2 \times 10^{-10}$ for ``low'' $m_T$, and $4.4 \times 10^{-10}$ for
$m_T \geq 300$ GeV, one order of magnitude above the SM prediction
$\mathrm{Br}(K_L \to \pi^0 \nu \bar \nu) = 2.4 \times 10^{-11}$. These rates
would be visible already at the E391 experiment at KEK, which aims at a
sensitivity of $3 \times 10^{-10}$, and up to $\sim 10^3$ events could be
collected at the KOPIO experiment approved for construction at BNL (for a
summary of the prospects on the rare decays $K^+ \to \pi^+ \nu \bar \nu$ and
$K_L \to \pi^0 \nu \bar \nu$ see for instance Ref.~\cite{papiro73}). The ratio
$\mathrm{Br}(K_L \to \pi^0 \nu \bar \nu)/
\mathrm{Br}(K^+ \to \pi^+ \nu \bar \nu)$ of the decay rates of the two kaon
``golden modes'', plotted in Fig.~\ref{fig:brk}, can be enhanced an order of
magnitude over the SM prediction $\sim 0.35$, and saturate the limit in
Eq.~(\ref{ec:n1}) for $m_T \geq 310$ GeV. This enhancement and a larger value of
$\mathrm{Br}(K^+ \to \pi^+ \nu \bar \nu)$ (compatible with experimental data)
lead to the maximum value
$\mathrm{Br}(K_L \to \pi^0 \nu \bar \nu) = 4.4 \times 10^{-10}$. On the other
hand, a strong suppression of this decay mode is possible, with values several
orders of magnitude below the SM prediction.

\begin{figure}[htb]
\begin{center}
\mbox{\epsfig{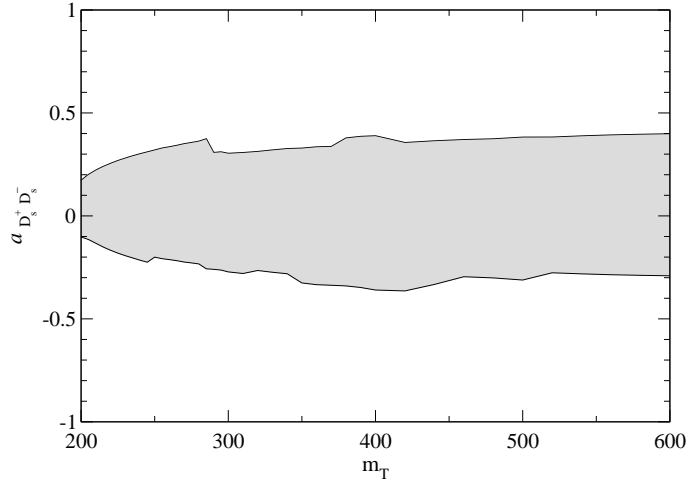}}
\end{center}
\caption{Range of variation of $a_{D_s^+ D_s^-}$ in Model I (shaded area).
\label{fig:aDsDs} }
\end{figure}

The mass difference in the $B_s$ system is predicted to be $\sim 18$ ps$^{-1}$
within the SM. The existing lower bound $|\delta m_{B_s}| \geq 13.1$ ps$^{-1}$
can be saturated in practically all the interval of $m_T$ studied. The ratio
$|\delta m_{B_s}/\delta m_{B_d}|$ has been proposed for a determination of
$|V_{ts}/V_{td}|$ \cite{papiro3}. Of course, this determination is strongly
model-dependent, because new physics may contribute to both mass differences.
This ratio equals 36 in the SM, and in Model I it may have values between the
experimental lower limit of 26.7 and 77. Finally, the asymmetry
$a_{D_s^+ D_s^-}$, which practically vanishes in the SM, provides a crucial test
of the phase structure of the CKM matrix. The non-unitarity of the $3 \times 3$
CKM submatrix and the presence of extra CP violating phases in Model I allow the
asymmetry $a_{D_s^+ D_s^-}$ to vary between $-0.4$ and $0.4$ for $m_T \geq 275$
GeV, as can be observed in Fig. \ref{fig:aDsDs}.

\subsection{Mixing with a down singlet}
In Model $\modii$ the mass of the new quark does not play an important r\^ole
in the constraints on the parameters of the model. The only dependence on $m_B$
appears in the $D^0$ mass difference (which at present does not imply any
restriction at least for masses up to 1 TeV), $b \to s \gamma$ (less restrictive
than $b \to s l^+ l^-$) and oblique parameters, which are less important than
$R_b$ and have no influence in practice. Agreement of the latter with experiment
requires that $|V_{tb}|$ is very close to unity, $|V_{tb}| \geq 0.998$. This is
indistinguishable from the SM prediction $|V_{tb}| = 0.999$, and forces $V_{td}$
and $V_{ts}$ to be within the SM range, $0.0059 \leq |V_{td}| \leq 0.013$,
$0.035\leq |V_{ts}| \leq 0.044$. The CKM matrix elements involving the new quark
are all small, $|V_{uB}| \leq 0.087$, $|V_{cB}| \leq 0.035$,
$|V_{tB}| \leq 0.041$, but noticeably they can be larger than $V_{ub}$.

\begin{figure}[htb]
\begin{center}
\mbox{\epsfig{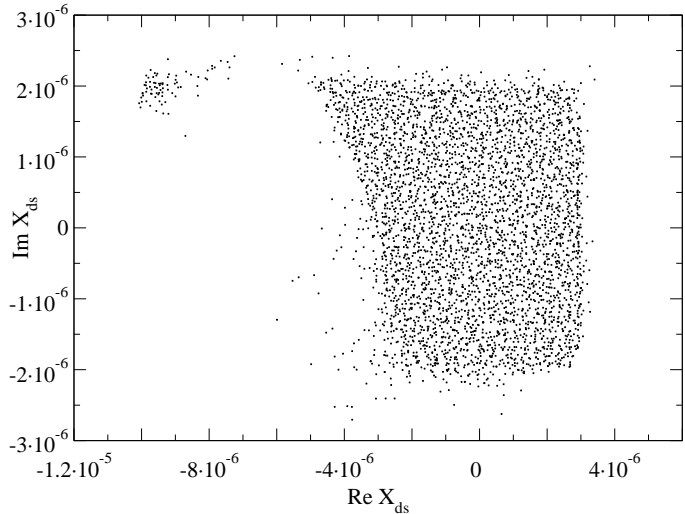}}
\end{center}
\caption{Allowed region for the real and imaginary parts of the FCN coupling
$X_{ds}$ in Model $\modii$.
\label{fig:xds} }
\end{figure}

FCN couplings between the light quarks are small (as required by low energy
observables), especially the coupling between the $d$ and $s$ quarks,
$|X_{ds}| \leq 1.0 \times 10^{-5}$. It makes sense to study
$\mathrm{Re}\,X_{ds}$ and $\mathrm{Im}\,X_{ds}$ separately, even though in
principle $X_{ds}$ is not a rephasing-invariant quantity. This is so because
Eq.~(\ref{ec:35}) assumes a CKM parameterisation with $V_{ud}^* V_{us}$ real.
This requirement eliminates the freedom to rephase $X_{ds}$ (up to a minus sign)
and enables to separate its real and imaginary parts meaningfully. The region of
allowed values for $X_{ds}$ is plotted in Fig.~\ref{fig:xds} for comparison with
other analyses in the literature \cite{papiro15a,papiro15b,papiro15c}. This
figure must be interpreted with care: the density of points is not associated to
any meaning of ``probability'', but it is simply an effect related to the random
generation and CKM parameterisation used to obtain the data points, and the
finiteness of the sample. The height of the allowed area is determined by the
$\varepsilon'/\varepsilon$ constraint, and the width by $K_L \to \mu^+ \mu^-$.
Comparing this plot with the ones in Refs.~\cite{papiro15a,papiro15c} we see
that the left part of the rectangle determined by $\varepsilon'/\varepsilon$ and
$K_L \to \mu^+ \mu^-$ is practically eliminated by the constraints from
$K^+ \to \pi^+ \nu \bar \nu$ and $\varepsilon$, except the upper left corner.
The height of the rectangle is also smaller, meaning that in our case the
requirement from $\varepsilon'/\varepsilon$ (using the prescription in
Appendix~\ref{sec:ap3}) is more stringent.

The upper bounds for $X_{db}$ and $X_{sb}$ found in our analysis are
$|X_{db}| \leq 1.1 \times 10^{-3}$, $|X_{sb}| \leq 1.1 \times 10^{-3}$. Plots
analogous to Fig.~\ref{fig:xds} are not meaningful for these parameters, because
there is a freedom to rephase the $b$ field and change arbitrarily the phases of
$X_{db}$ and $X_{sb}$. The only meaningful bounds are hence the limits on their
moduli. The FCN coupling $X_{bB}$ is not so limited by low energy measurements,
and can reach $0.041$.

Despite these restrictions on $X_{ds}$, $X_{db}$, $X_{sb}$ and the fact that CKM
matrix elements involving the known quarks must be within the SM range, the
presence of tree-level FCN couplings has sizeable effects on some low energy
observables, of the same magnitude as in Model I. A decay rate
$\mathrm{Br}(K_L \to \pi^0 \nu \bar \nu) = 1.6 \times 10^{-10}$ can be achieved,
and the  ratio $\mathrm{Br}(K_L \to \pi^0 \nu \bar \nu)/
\mathrm{Br}(K^+ \to \pi^+ \nu \bar \nu)$ can equal 4.34. The lower limit on
$|\delta m_{B_s}|$ can be saturated, and the ratio
$|\delta m_{B_s}/\delta m_{B_d}|$ can be up to $67$. On the other hand, the 
asymmetry $a_{D_s^+ D_s^-}$ takes values between $-3 \times
10^{-3}$ and $0.11$. This upper limit is a factor of $2-3$ larger than the SM
prediction.

\subsection{Mixing with two singlets}
Mixing with more than one singlet lets two quarks of the same charge, for
instance the $d$ and $s$ quarks, mix significantly with exotic quarks without
necessarily generating a FCN coupling $X_{ds}$ between them, in virtue of
Eqs.~\ref{ec:6}. This allows a better fit to the measured CKM matrix elements
and $u$, $d$ diagonal couplings to the $Z$ boson, especially in Model $\modii$.
In this model the global fit can be considerably better than in the SM, for
instance with the CKM matrix
\begin{eqnarray}
|V| & = & \left( \mbox{\begin{tabular}{lllll}
0.9742 & 0.2187 & 0.0037 & 0.0325 & 0.0442 \\
0.2183 & 0.9750 & 0.0401 & 0.0076 & 0.0097 \\
0.0074 & 0.0396 & 0.9992 & 0.0061 & 0.0036 \\
0.0539 & 0.0001 & 0.0028 & 0.7370 & 0.6737 \\
0.0160 & 0.0002 & 0.0066 & 0.6750 & 0.7376
\end{tabular}} \right) \nonumber \,, \\
\arg V & = & \left( \mbox{\begin{tabular}{lllll}
0.00 & 0.00 & 3.09 & 0.00 & 0.00 \\
5.63 & 2.49 & 0.28 & 5.66 & 5.61 \\
1.79 & 5.36 & 0.00 & 0.02 & 2.92 \\
0.69 & 2.56 & 0.00 & 3.60 & 4.02 \\
3.96 & 2.70 & 0.00 & 2.91 & 0.19
\end{tabular}} \right) \,.
\end{eqnarray}
The actual masses of the two extra down quarks are not very relevant, and have
been taken as 200 and 400 GeV in the calculation. In this example the $\chi^2$
of the six measured CKM matrix elements is 1.14, while in the SM best fit it is
4.77. The parameters describing the $Zuu$ and $Zdd$ couplings are
$g_L^2 = 0.3024$, $\theta_L = 2.4612$, $C_{1d} = 0.3398$,
$\tilde C_2 = -.0492$, and the others unchanged with respect to their SM values.
The $\chi^2$ of these parameters is 7.73, improving the SM value of 10.5. The
agreement of the rest of the observables with experiment is equal or better
than within the SM, as can be seen in Table~\ref{tab:10a} (the experimental
results for $\varepsilon$ and $\delta m_B$ can be accommodated with slightly
larger $\hat B$ parameters). This example of ``best fit'' matrix gives the
predictions $|X_{bB}| = 0.006$, $|X_{bB'}| = 0.004$.
The result $\mathrm{Br}(K_L \to \pi^0 \nu \bar \nu) = 2.0 \times 10^{-10}$ is
very similar to the SM case, but other examples with a little worse $\chi^2$
can be found, having enhancements (or suppressions) of this rate by factors up
to three. These examples show explicitly that new physics effects are
not in contradiction with good agreement with experimental data, although our
restrictive criteria for agreement with experiment at the beginning of this
Section already made it apparent.

\begin{table}[htb]
\begin{center}
\begin{tabular}{cc}
\hline
\hline
Observable & Value \\
\hline
$R_b$ & $0.21590$ \\
$R_c$ & $0.1724$ \\
$A_\mathrm{FB}^{0,b}$ & $0.1039$ \\
$A_\mathrm{FB}^{0,c}$ & $0.0744$ \\
$\mathcal{A}_b$ & $0.935$ \\
$\mathcal{A}_c$ & $0.669$ \\
$\varepsilon$ & $2.08 \times 10^{-3}$ \\
$|\delta m_B|$ & $0.45$ \\
$|\delta m_{B_s}|$ & $17.6$ \\
$a_{\psi K_S}$ & $0.74$ \\
$\mathrm{Br}(K^+ \to \pi^+ \nu \bar \nu)$ & $6.0 \times 10^{-11}$ \\
$\mathrm{Br}(K_L \to \mu^+ \mu^-)_\mathrm{SD}$ & $6.3 \times 10^{-10}$ \\
$\mathrm{Br}(b \to s \gamma)$ & $3.35 \times 10^{-3}$ \\
$\mathrm{Br}(b \to s e^+ e^-)$ & $7.3 \times 10^{-6}$ \\
$\mathrm{Br}(b \to s \mu^+ \mu^-)$ & $5.0 \times 10^{-6}$ \\
$\varepsilon'/\varepsilon$ & $1.6 \times 10^{-3}$ \\
\hline
\hline
\end{tabular}
\end{center}
\caption{Values of some observables for the ``best fit'' matrix in Model
$\modii$ with two extra singlets. The mass differences are in ps$^{-1}$.
\label{tab:10a}}
\end{table}

Finally, we have also noticed that the predictions for the parameters and
observables under study do not change appreciably neither in Model I nor in
Model $\modii$ when we allow mixing with more than one singlet of the same
charge.

\section{Conclusions}
\label{sec:9}
The aim of this paper has been to investigate how the existence of a new quark
singlet may change many predictions of the SM while keeping agreement with
present experimental data. In Model I the mixing with a $Q=2/3$ singlet might
lead to huge departures from the SM expectation for the CKM matrix elements
$V_{td}$, $V_{ts}$, $V_{tb}$ and the diagonal coupling $Ztt$. Additionally,
observable FCN couplings $Ztu$ and $Ztc$ may appear. These effects depend on the
mass of the new quark, as has been shown in Figs.~\ref{fig:vtb}-\ref{fig:brk}. 
For $m_T \sim 200$ GeV the new quark might effectively replace the top in
reproducing the experimental observables in $K$ and $B$ physics, allowing for
values of $V_{td}$ and $V_{ts}$ very different from the SM predictions. On the
other hand, for larger $m_T$ the leading contributions to $K$ and $B$
observables are the SM ones, with possible new contributions from the new quark.
This effect can be clearly appreciated in Figs.~\ref{fig:vts}--\ref{fig:vr},
where it is also apparent how important a direct determination of $V_{td}$ and
$V_{ts}$ would be. Unfortunately, the difficulty in tagging light quark jets at
Tevatron and LHC  makes these measurements very hard, if not impossible. Any
experimental progress in this direction would be most welcome.

The mixing of the top with the new quark results in values of $V_{tb}$ and the
$Ztt$ coupling parameter $X_{tt}$ significantly smaller than one. These
deviations from unity would be observable at LHC \cite{papiro9a} and TESLA,
respectively. For larger $m_T$, $|V_{tb}|$ and $|X_{tt}|$ must be closer to
unity, as can be seen in Figs.~\ref{fig:vtb} and \ref{fig:xtt}. However, the
decrease in $X_{tt}$ would be visible at TESLA even for $m_T = 600$ GeV. The FCN
couplings $Ztu$ and $Ztc$ could also be observed at LHC for a wide range of
$m_T$ \cite{papiro66,papiro67,papiro68}. 

The effects of top mixing are not limited to large colliders. Indeed, the
observables in $K$ and $B$ physics studied here provide an example where these
effects do not disappear when the mass of the new quark is large. We have shown
that the predictions for the decay $K_L \to \pi^0 \nu \bar \nu$, the
$\delta m_{B_s}$ mass difference and the CP asymmetry $a_{D_s^+ D_s^-}$ can be
very different from the SM expectations, and  effects of new physics could be
observed in experiments under way or planned. These predictions for Model I are
collected in Table~\ref{tab:10}. Before LHC operation, indirect evidences of new
physics could appear in the measurement of CP asymmetries at $B$ factories. A
good candidate is the asymmetry $a_{D_s^+ D_s^-}$ discussed here, but many other
observables and CP asymmetries are worth analysing. If no new physics is
observed, further constraints could be placed on CP violating phases.

\begin{table}[htb]
\begin{center}
\begin{tabular}{cccc}
\hline
\hline
Quantity & ~~~~ & \multicolumn{2}{c}{Range} \\
\hline
$|V_{tb}|$ & & $0.58$ & $1$ \\
$|V_{td}|$ & & $4 \times 10^{-5}$ & $0.044$ \\
$|V_{ts}|$ & & $0.002$ & $0.06$ \\
$|V_{td}/V_{ts}|$ & & $6 \times 10^{-4}$ & $2.9$ \\
$|V_{Td}|$ & & $0$ & $0.052$ \\
$|V_{Ts}|$ & & $0$ & $0.063$ \\
$|V_{Tb}|$ & & $0$ & $0.81$ \\
$|X_{tt}|$ & & $0.34$ & $1$ \\
$|X_{ut}|$ & & $0$ & $0.038$ \\
$|X_{ct}|$ & & $0$ & $0.036$ \\
$|X_{tT}|$ & & $0$ & $0.5$ \\
$\mathrm{Br}(K_L \to \pi^0 \nu \bar \nu)$ & & $\sim 0$ &
  $4.4 \times 10^{-10}$ \\
$\frac{\mathrm{Br}(K_L \to \pi^0 \nu \bar \nu)}{
  \mathrm{Br}(K^+ \to \pi^+ \nu \bar \nu)}$ & & $\sim 0$ & $4.35$ \\
$|\delta m_{B_s}/\delta m_B|$ & & $26.7$ & $77$ \\
$a_{D_s^+ D_s^-}$ & & $-0.4$ & $0.4$ \\[0.1cm]
\hline
\hline
\end{tabular}
\caption{Summary of the predictions for Model I.
\label{tab:10}}
\end{center}
\end{table}

In Model $\modii$ the effects of the new $Q=-1/3$ singlet on CKM matrix elements
are negligible and FCN couplings between known quarks are very constrained by
experimental data. However, the predictions for meson observables, summarised
in Table~\ref{tab:11}, are rather alike. In addition, we have shown how the
mixing with two singlets can improve the agreement with the experimental
determination of CKM matrix elements and $Zuu$, $Zdd$ couplings. This can be
done keeping similar and in some cases better agreement with electroweak
precision data and $K$ and $B$ physics observables.

\begin{table}[htb]
\begin{center}
\begin{tabular}{cccc}
\hline
\hline
Quantity & ~~~~ & \multicolumn{2}{c}{Range} \\
\hline
$|V_{tb}|$ & & $0.998$ & $1$ \\
$|V_{td}|$ & & $0.0059$ & $0.013$ \\
$|V_{ts}|$ & & $0.035$ & $0.044$ \\
$|V_{uB}|$ & & $0$ & $0.087$ \\
$|V_{cB}|$ & & $0$ & $0.035$ \\
$|V_{tB}|$ & & $0$ & $0.041$ \\
$|X_{ds}|$ & & $0$ & $1.0 \times 10^{-5}$ \\
$\mathrm{Re}\, X_{ds}$ & & $-1.0 \times 10^{-5}$ & $3.4 \times 10^{-6}$ \\
$\mathrm{Im}\, X_{ds}$ & & $-2.7 \times 10^{-6}$ & $2.4 \times 10^{-6}$ \\
$|X_{db}|$ & & $0$ & $1.1 \times 10^{-3}$ \\
$|X_{sb}|$ & & $0$ & $1.1 \times 10^{-3}$ \\
$|X_{bB}|$ & & $0$ & $0.041$ \\
$\mathrm{Br}(K_L \to \pi^0 \nu \bar \nu)$ & & $\sim 0$ &
  $1.6 \times 10^{-10}$ \\
$\frac{\mathrm{Br}(K_L \to \pi^0 \nu \bar \nu)}{
  \mathrm{Br}(K^+ \to \pi^+ \nu \bar \nu)}$ & & $\sim 0$ & $4.34$ \\
$|\delta m_{B_s}/\delta m_B|$ & & $26.7$ & $67$ \\
$a_{D_s^+ D_s^-}$ & & $-3 \times 10^{-3}$ & $0.11$ \\[0.1cm]
\hline
\hline
\end{tabular}
\caption{Summary of the predictions for Model $\modii$.
\label{tab:11}}
\end{center}
\end{table}

All the effects of mixing with singlets described are significant, but of
course the decisive evidence would be the discovery of a new quark, which might
happen at LHC or even at Tevatron, provided it exists and it is light enough.
In this case, the pattern of new physics effects would allow to uncover its
nature. Conversely, the non-observation of a new quark would be very important
as well. If no new quark is found at LHC, the indirect constraints on CKM matrix
elements and nonstandard contributions to meson physics would considerably
improve.

\appendix
\section{Common input parameters}
\label{sec:ap1}
Unless otherwise specified, experimental data used throughout the paper
are taken from Refs.~\cite{papiro1,papiro3}. We use the results in
Ref.~\cite{papiroap1} to convert the pole masses $m_i$ to the
$\overline \mathrm{MS}$ scheme and to perform the running to the scale $M_Z$.
The results are in Table~\ref{tab:ap1}. For $u$, $d$, $s$ we quote the
$\overline \mathrm{MS}$ masses at 2 GeV instead of the pole masses. The numbers
between brackets are not directly used in the calculations.

\begin{table}[htb]
\begin{center}
\begin{tabular}{cccc}
\hline
\hline
& $m_i$ & $\overline m_i(m_i)$ & $\overline m_i(M_Z)$ \\
\hline
$m_u$ & $(0.003)$ & $-$ & $0.0016$ \\
$m_d$ & $(0.006)$ & $-$ & $0.0033$ \\
$m_c$ & $1.5$ & $1.22$ & $0.68$  \\
$m_s$ & $0.12$ & $-$ & $0.067$ \\
$m_t$ & 174.3 & 164.6 & 175.6 \\
$m_b$ & $4.7$ & $4.12$ & $2.9$ \\
\hline
\hline
\end{tabular}
\caption{Quark masses (in GeV) used in the evaluations. The uncertainty in $m_t$
is taken as $\pm 5.1$ GeV. For $u$, $d$, $s$ we write the
$\overline \mathrm{MS}$ masses $\overline m_i(2~\mathrm{GeV})$ instead of the
pole masses.
\label{tab:ap1}}
\end{center}
\end{table}

The running masses $\overline m_c(\overline m_c) = 1.28$,
$\overline m_b(\overline m_b) = 4.19$ are also needed. The lepton pole masses
are $m_e = 0.511$ MeV, $m_\mu = 0.105$ and $m_\tau = 1.777$ GeV. We take
$M_Z = 91.1874$, $\Gamma_Z = 2.4963$, $M_W = 80.398$ and $M_H = 115$ GeV.
The electromagnetic and strong coupling constants at the scale $M_Z$ are
$\alpha = 1/128.878$, $\alpha_s = 0.118$. The sine of the weak angle in the
$\overline \mathrm{MS}$ scheme is $s_Z^2 = 0.23113$.

The CKM matrix used in the context of the SM is obtained by a fit to the six
measured moduli in Table~\ref{tab:0}, and is determined by $|V_{us}| = 0.2224$,
$|V_{ub}| = 0.00362$, $|V_{cb}| = 0.0402$, and the rest of the elements obtained
using $3 \times 3$ unitarity. The phase $\delta$ in the standard
parameterisation \cite{papiro3} is determined performing a fit to 
$\varepsilon$, $\varepsilon'/\varepsilon$, $a_{\psi K_S}$ and $|\delta m_B|$
with the rest of parameters quoted, and the result $\delta = 1.014$ is very
similar to the one obtained in the fit in Ref.~\cite{papiro3}.

\section{Inami-Lim functions}
\label{sec:ap2}
In this Appendix we collect the Inami-Lim functions used in Section~\ref{sec:5}.
The box functions $F$ and $S_0$ appear in meson oscillations. $D_0'$, $E_0$ and
$E_0'$ are related to photon and gluon penguins. The functions $X_0$ and $Y_0$
are gauge-invariant combinations of the box function $B_0$ and the $Z$ penguin
function $C_0$, $X_0 = C_0-4 B_0$, $Y_0 = C_0-B_0$. The  function $Z_0$ is a
gauge-invariant combination of photon and $Z$ penguins. Their expressions read
\cite{papiro27,papiro29}
\begin{eqnarray}
E_0(x_i) & = & -\frac{2}{3} \log x_i
+ \frac{x_i(18-11 x_i-x_i^2)}{12(1-x_i)^3}
+ \frac{x_i^2 (15-16 x_i + 4 x_i^2)}{6 (1-x_i)^4} \log x_i \,,
\\
D_0'(x_i) & = & -\frac{8 x_i^3+5 x_i^2-7 x_i}{12(1-x_i)^3}
+ \frac{-3 x_i^3+2 x_i^2}{2(1-x_i)^4} \log x_i \,,
\\
E_0'(x_i) & = & -\frac{x_i^3-5 x_i^2-2 x_i}{4(1-x_i)^3}
+ \frac{3 x_i^2}{2(1-x_i)^4} \log x_i \,,
\\
F(x_i,x_j) & = & \frac{4-7 x_i x_j}{4 (1-x_i) (1-x_j)}
+ \frac{4-8 x_j+x_i x_j}{4 (1-x_i)^2 (x_i-x_j)} x_i^2 \log x_i \nonumber \\
& & + \frac{4-8 x_i+x_i x_j}{4 (1-x_j)^2 (x_j-x_i)} x_j^2 \log x_j \,,
\\
S_0(x_i,x_j) & = & -\frac{3 x_i x_j}{4 (x_i-1) (x_j-1)}
+ \frac{x_i x_j \left( x_i^2-8 x_i+4 \right)}{4 (x_i-1)^2 (x_i-x_j)} \log x_i
\nonumber \\
& & + \frac{x_i x_j \left( x_j^2-8 x_j+4 \right)}{4 (x_j-1)^2 (x_j-x_i)} 
\log x_j \,, 
\\
S_0(x_i) & = & \frac{4 x_i-11 x_i^2+x_i^3}{4 (1-x_i)^2} 
- \frac{3 x_i^3}{2 (1-x_i)^3} \log x_i \,,
\\
Z_0(x_i) & = & -\frac{1}{9} \log x_i
+ \frac{18 x_i^4-163 x_i^3+259 x_i^2 -108 x_i}{144 (x_i-1)^3} \nonumber \\
& & + \frac{32 x_i^4-38 x_i^3-15 x_i^2+18 x_i}{72 (x_i-1)^4} \log x_i \,,
\\
B_0(x_i) & = & \frac{1}{4} \left[ \frac{x_i}{1-x_i}
+ \frac{x_i}{(x_i-1)^2} \log x_i \right] \,,
\\
C_0(x_i) & = & \frac{x_i}{8} \left[ \frac{x_i-6}{x_i-1}
+ \frac{3 x_i+2}{(x_i-1)^2} \log x_i \right] \,,
\\
X_0(x_i) & = & \frac{x_i}{8} \left[ \frac{x_i+2}{x_i-1}
+ \frac{3 x_i-6}{(x_i-1)^2} \log x_i \right] \,,
\\
Y_0(x_i) & = & \frac{x_i}{8} \left[ \frac{x_i-4}{x_i-1}
+ \frac{3 x_i}{(x_i-1)^2} \log x_i \right] \,.
\end{eqnarray}
The functions appearing in the $Z$ FCNC penguins involved in the calculation
of $b \to s \gamma$ are
\cite{papiro45}
\begin{eqnarray}
\xi_s^Z & = & \frac{1}{54}(-3+2 s_W^2) \,,  \\
\xi_b^Z & = & \frac{1}{54}(-3-4 s_W^2) \,,  \\
\xi_B^Z(y_B) & = & -\frac{8-30 y_B + 9 y_B^2 - 5 y_B^3}{144 (1-y_B)^3}
 + \frac{y_B^2}{8(1-y_B)^4} \log y_B \,, \\
\xi_B^H(w_B) & = & -\frac{16 w_B-29 w_B^2+7 w_B^3}{144(1-w_B)^3}
 + \frac{-2w_B+3 w_B^2}{24(1-w_B)^4} \log w_B \,,
\end{eqnarray}
where we have approximated $y_s=0$, $y_b=0$ and $m_s/m_b=0$.

\section{Statistical analysis of observables with theoretical uncertainty}
\label{sec:ap3}
The most common situation when comparing a theoretical prediction $x_{t}$ with
an experimental measurement $x_e$ is that the uncertainty in the former can be
ignored. This does not happen for some observables analysed in this article,
which are subject to low energy QCD uncertainties. For example, if we have for
$\varepsilon$ $x_e = (2.282 \pm 0.017) \times 10^{-3}$ and
$x_t = (2.42 \pm 0.42) \times 10^{-3}$, how many standard deviations is $x_t$
from $x_e$? To answer naively that it is at $8.1\sigma$ is clearly wrong, and
the comparison between both should weigh in some way the error on $x_t$. Here we
explain how we obtain in such cases a reasonable estimate of the agreement
between the theoretical and experimental data.

Let us recall how $x_e$ and $x_t$ are compared when the former has a Gaussian
distribution with mean $\mu_e$ and standard deviation $\sigma_e$ and $x_t$ is
error-free and equals $\mu_t$ (see for instance Ref.~\cite{papiroap3}). The
$\chi^2$ value is defined as
\begin{equation}
\chi^2 = \left( \frac{\mu_e-\mu_t}{\sigma_e} \right)^2 \,,
\label{ec:ap3-1}
\end{equation}
and from it the $P$ number is computed as
\begin{equation}
P = \int_{\chi^2}^\infty f(z;1) \,dz,
\label{ec:ap3-2}
\end{equation}
where $f(z;n)$ is the $\chi^2$ distribution function for $n$ degrees of freedom,
\begin{equation}
f(z;n) = \frac{z^{n/2-1} e^{-z/2}}{2^{n/2} \Gamma(n/2)} \,.
\label{ec:ap3-3}
\end{equation}
The $P$ value is the probability to obtain experimentally a $\chi^2$ equal or
worse than the actual one, that is, a result equal or less compatible with the
theory. Performing the integral in Eq.~(\ref{ec:ap3-2}),
\begin{equation}
P = 1-\mathrm{erf} \, \sqrt{\frac{\chi^2}{2}} = 1-\mathrm{erf} \,
\frac{|\mu_e-\mu_t|}{\sqrt 2 \,\sigma_e} \,,
\label{ec:ap3-4}
\end{equation}
with $\mathrm{erf}$ the well-known error function. The probability to obtain an
equal or better result is $1-P$. For instance, with $|\mu_e-\mu_t| = \sigma_e$
we have $1-P=\mathrm{erf} (1/\sqrt 2) = 0.68$, corresponding to one Gaussian
standard deviation, as it obviously must be.

When $x_t$ is not considered as a fixed quantity $\mu_t$ but has some
distribution function $g(x_t)$ (that may be Gaussian or may not), we use the
probability law $P(A)=\sum_i P(A|B_i) P(B_i)$, with $\sum_i P(B_i)=1$, to
convolute the $x_t$-dependent $P$ number with $g$:
\begin{equation}
P  =  \int_{-\infty}^{+\infty} P|_{\mu_t \to x_t} g(x_t) \,dx_t 
= 1 - \int_{-\infty}^{+\infty} \mathrm{erf} \,
\frac{|\mu_e-x_t|}{\sqrt 2 \,\sigma_e} g(x_t) \,dx_t \,.
\label{ec:ap3-5}
\end{equation}
The assumption that $x_t = \mu_t$ without error can be translated to
Eq.~(\ref{ec:ap3-5}) choosing the ``distribution function''
$g(x_t) = \delta(x_t-\mu_t)$, in which case we recover Eq.~(\ref{ec:ap3-4}).

An adequate (but not unique) choice of the function $g(x_t)$ may be a Gaussian.
One source of systematic uncertainties is often due to the
input parameters involved in the theorical calculation ($m_t$,
$\alpha_s$, CKM mixing angles, etc.), whose experimental values are given by a
Gaussian
distribution. It is then likely that the distribution function $g(x_t)$
(where $x_t$ is also a function of its input parameters)
has a maximum at $\mu_t$ and falls quickly for increasing $|x_t-\mu_t|$.
This feature can be implemented in a simple way by choosing $g(x_t)$ as
Gaussian, and we expect that the results are not very sensitive to the precise
shape of the function $g(x_t)$.

Let us then assume that $g(x_t)$ is a Gaussian with mean $\mu_t$ and standard
deviation $\sigma_t$. Intuitively, we expect that if $\sigma_t \ll \sigma_e$,
Eq.~(\ref{ec:ap3-5}) should reduce to Eq.~(\ref{ec:ap3-4}). This is easy to
show. Writing the explicit form of $g(x_t)$,
\begin{equation}
P = 1 - \int_{-\infty}^{+\infty}
\mathrm{erf} \, \frac{|\mu_e-x_t|}{\sqrt 2 \,\sigma_e} \;
\frac{e^{-\frac{(x_t-\mu_t)^2}{2 \sigma_t^2}}}{\sqrt{2 \pi} \sigma_t}
\,dx_t \,.
\label{ec:ap3-7}
\end{equation}
The limits of this integral can be taken as $\mu_t-n\, \sigma_t$,
$\mu_t + n\, \sigma_t$, with $n \geq 4$. The integral is negligible out of these
limits due to the exponential (the error function takes values between 0 and 1).
Changing variables to $\Delta_t = x_t - \mu_t$, we observe that
$|\Delta_t| \ll \sigma_e$ under the assumption that $\sigma_t \ll \sigma_e$.
Expanding the error function in a Taylor series to order $\Delta_t$, the
integral can be done analytically,
\begin{equation}
P = 1- \mathrm{erf}\,\frac{n}{\sqrt 2} \; \mathrm{erf} \,
\frac{|\mu_e-\mu_t|}{\sqrt 2 \,\sigma_e} \,.
\label{ec:ap3-8}
\end{equation}
For $n \geq 4$, $\mathrm{erf}\,n/\sqrt 2 \simeq 1$ to an excellent approximation
and we obtain Eq.~(\ref{ec:ap3-4}), as we wanted to prove.

Results for $P$ values can be expressed in a more intuitive form as standard
``number of sigma'' $n_\sigma$ inverting Eq.~(\ref{ec:ap3-4}),
\begin{equation}
n_\sigma = \sqrt 2 \, \mathrm{erf}^{-1}(1-P) \,,
\label{ec:ap3-6}
\end{equation}
with $\mathrm{erf}^{-1}$ the inverse of the error function. However, this
$n_\sigma$ does not retain the geometrical interpretation of the distance
between $\mu_e$ and $\mu_t$ in units of $\sigma_e$ that has when
$\sigma_t \sim 0$.

We apply this procedure to the example at the beginning of this Appendix,
with $x_e = (2.282 \pm 0.017) \times 10^{-3}$ and
$x_t = (2.42 \pm 0.42) \times 10^{-3}$.
Assuming for simplicity that the distribution of $x_t$ is Gaussian, we obtain
the much more
reasonable result of $n_\sigma = 2.25$. This number must be compared
with $n_\sigma=8.1$, obtained without taking into account the
theoretical error, {\em i. e.} calculating naively $(\mu_t-\mu_e)/\sigma_e$.
The use of the theoretical error in the statistical comparison mitigates the
discrepancy and implements numerically, in a simple but effective way,
what one would intuitively expect in this case. The result $n_\sigma = 2.25$
reflects the fact that the theoretical and experimental values can have a
good agreement if $x_t$ is smaller than its predicted value
$\mu_t = 2.42 \times 10^{-3}$ but a bad one
if $x_t$ is larger, what is also possible because the theoretical error is
of either sign. The
prescription presented here has also one very gratifying property: if we change
$\mu_t \to \mu'_t = \mu_t+\delta \mu_t$, with $\delta \mu_t \ll \sigma_t$, the
$P$ value is hardly affected. For $\mu'_t = 2.35 \times 10^{-3}$, $n_\sigma$
changes only to $2.23$, while the pull calculated naively decreases to $4$.

This construction can be generalised when not all the range of variation of
$x_e$, $x_t$ is physically allowed. We write without proof the expression for
$P$ in this case. Assuming that the physical region is $x_e \geq 0$,
$x_t \geq 0$,
\begin{eqnarray}
P & = & 1- \left[ 1+\mathrm{erf}\, \frac{\mu_e}{\sqrt 2 \sigma_e} \right]^{-1}
\nonumber \\[0.1cm]
& & \times \int_0^\infty \left(
\mathrm{erf} \, \frac{|\mu_e-x_t|}{\sqrt 2 \,\sigma_e} +
\mathrm{erf} \, \frac{\mathrm{max}\{\mu_e,|\mu_e-x_t|\} }{\sqrt 2 \,\sigma_e}
\right) \, g(x_t) \,dx_t \,.
\label{ec:ap3-9}
\end{eqnarray}

Finally, notice that the expressions in Eqs.~(\ref{ec:ap3-5},\ref{ec:ap3-9}) for
the $P$ number are not symmetric under the interchange of theoretical and
experimental data, even if $g(\mu_t)$ is Gaussian. This reflects the fact that
$P(\mathrm{data}|\mathrm{theory}) \neq P(\mathrm{theory}|\mathrm{data})$, but
they are related by Bayes' theorem.

\vspace{1cm}
\noindent
{\Large \bf Acknowledgements}

\vspace{0.4cm} \noindent
I thank F. del \'Aguila, R. Gonz\'alez Felipe, F. Joaquim, J. Prades,
J. Santiago and J. P. Silva for useful comments. I also thank J. P. Silva,
F. del \'Aguila, A. Teixeira and G. C. Branco for reading the manuscript.
This work has been supported by the European Community's Human Potential
Programme under contract HTRN--CT--2000--00149 Physics at Colliders and by FCT
through project CERN/FIS/43793/2001.

\end{document}